\documentclass[pra,aps,twocolumn,superscriptaddress,notitlepage,10pt]{revtex4-2}
\usepackage{amssymb,amsfonts,amsmath,amstext,graphicx,hyperref}
\usepackage{tikz}
\usepackage{graphicx}
\usepackage[normalem]{ulem}
\usepackage{tabularx}

\def\BraVert{\egroup\,\mid\,\bgroup}

\usepackage{braket}
\hypersetup{
	colorlinks=true,
	linkcolor=blue,
	filecolor=magenta,      
	urlcolor=cyan,
}

\renewcommand{\paragraph}[1]{\textit{#1}}

\begin{document}
\title{Generation of entanglement and non-stationary states via competing coherent and incoherent bosonic hopping}

\author{Parvinder Solanki}
\email{parvinder@mnf.uni-tuebingen.de}
\thanks{Part of this work was done at {Department of Physics, University of Basel, Klingelbergstrasse 82, CH-4056 Basel, Switzerland}.}
\affiliation{Institut f\"ur Theoretische Physik and Center for Integrated Quantum Science and Technology, Universit\"at T\"ubingen, Auf der Morgenstelle 14, 72076 T\"ubingen, Germany}
\author{Albert Cabot}
\affiliation{Institut f\"ur Theoretische Physik and Center for Integrated Quantum Science and Technology, Universit\"at T\"ubingen, Auf der Morgenstelle 14, 72076 T\"ubingen, Germany}
\author{Matteo Brunelli}
\affiliation{JEIP, UAR 3573 CNRS, Coll\`{e}ge de France, PSL Research University, 11 Place Marcelin Berthelot, 75321 Paris Cedex 05, France} 
\author{Federico Carollo}
\affiliation{Centre for Fluid and Complex Systems, Coventry University, Coventry, CV1 2TT, United Kingdom}
\author{Christoph Bruder}
\affiliation{Department of Physics, University of Basel, Klingelbergstrasse 82, CH-4056 Basel, Switzerland}
\author{Igor Lesanovsky}
\affiliation{Institut f\"ur Theoretische Physik and Center for Integrated Quantum Science and Technology, Universit\"at T\"ubingen, Auf der Morgenstelle 14, 72076 T\"ubingen, Germany}
\affiliation{School of Physics and Astronomy and Centre for the Mathematics and Theoretical Physics of Quantum Non-Equilibrium Systems, University of Nottingham, Nottingham, NG7 2RD, United Kingdom}

\date{\today}

\begin{abstract}

Incoherent stochastic processes added to unitary dynamics are typically deemed detrimental since they are expected to diminish quantum features such as superposition and entanglement. Instead of exhibiting energy-conserving persistent coherent motion, the dynamics of such open systems feature, in most cases, a steady state, which is approached in the long-time limit from all initial conditions. 
This can, in fact, be advantageous as it offers a mechanism for the creation of robust quantum correlations on demand without the need for fine-tuning. Here, we show this for a system consisting of two coherently coupled bosonic modes, which is a paradigmatic scenario for the realization of quantum resources such as squeezed entangled states. Rather counterintuitively, the mere addition of incoherent hopping, which results in a statistical coupling between the bosonic modes, leads to steady states with robust quantum entanglement and enables the emergence of persistent coherent non-stationary behavior.   
\end{abstract}

\maketitle

\paragraph{Introduction.---}
A central challenge of contemporary quantum physics is to uncover the emergent collective behavior of many-body systems and to harness it in technological applications \cite{sorensen2001many,fazio2024manybodyopenquantumsystems,korbicz2005spin}. A prime example is a Bose-Einstein condensate (BEC) \cite{griffin1996bose}, which is a state of matter that features coherence at a macroscopic level. Coupling several BECs and introducing interactions allows for the generation of a whole host of collective phenomena \cite{hofer2012superfluid,lode2017fragmented,hofer2012superfluid,PhysRevLett.120.215301,autti2021ac,kessler2021observation}, which, at a theoretical level, are captured by the Bose-Hubbard Hamiltonian. The most elementary instance, already displaying a  surprisingly rich dynamics, is a BEC in a double well \cite{PhysRevLett.79.4950}. In this setting,  the competition between coherent particle hopping and interactions  generates correlated spin-squeezed states \cite{orzel2001squeezed,esteve2008squeezing}, which have been used as a resource for quantum-enhanced metrology. This example is representative of a general class of protocols that produce quantum resources from the unitary time evolution of a carefully prepared initial state \cite{orzel2001squeezed,esteve2008squeezing}.

\begin{figure}[t!]
    \centering
    \includegraphics[width=\linewidth]{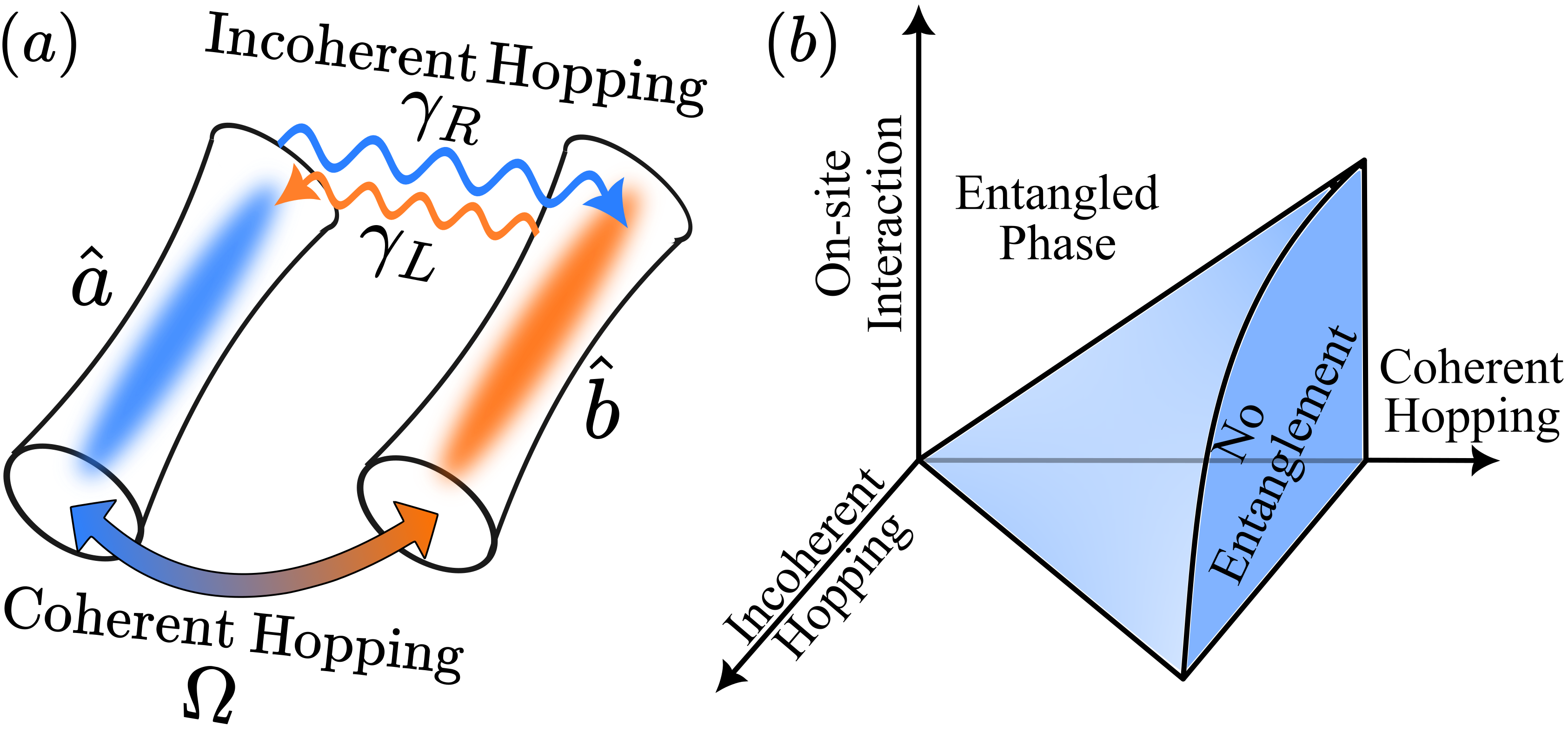}
    \caption{\textbf{Model and entanglement properties.} $(a)$ Illustration of a Bose-Hubbard dimer (BHD) showing coherent ($\Omega$) and incoherent hopping ($\gamma_{R,L}$), where each mode can also be subject to on-site interactions (not shown here) as described by Eq.~(\ref{eq:BH_themal}). The rate $\gamma_{R}$ ($\gamma_{L}$) parametrizes the incoherent hopping strength to the right (left). (b) Sketch of the steady-state entanglement as a function of parameters. Notably, stronger incoherent hopping favors entanglement generation.}
    \label{fig:intro}
\end{figure}

In this work, we show how adding a simple incoherent process to a unitary evolution can offer a route towards creating robust quantum resources. Robustness arises here from the fact that the resulting dissipative dynamics features a long-time steady state, which is reached without the need for specifically tuned initial conditions or timing requirements. In particular, we consider a system of two bosonic modes realized within a double-well BEC or through coupled cavities, also known as Bose-Hubbard dimer (BHD) [see the sketch in Fig.~\ref{fig:intro}(a)]. Tuning interactions, coherent and incoherent hopping strength, 
allows for controlling quantum correlations between the two modes.
Interestingly, larger incoherent hopping rates favor the generation of entanglement, as illustrated in Fig.~\ref{fig:intro}(b). Moreover, in certain parameter regimes, the system can even feature non-stationary states in the long-time limit, entering a so-called time-crystal phase in which the time-translation symmetry of the underlying dynamical generator is broken \cite{iemini2018boundary,tucker2018shattered,buca2019non,Booker_2020,PhysRevLett.123.260401,zhu2019dicke,lledo2019driven,pizzi2019periodn,pizzi2021bistability,seibold2020dissipative,prazeres2021boundary,piccitto2021symmetries,carollo2022exact,krishna2022measurement,solanki2022role,seeding2022michal,hurtado2020raretimecrystal,alaeian2022exact,PhysRevLett.127.133601,PhysRevA.107.L010201,iemini2024sectors,cabot2023nonequilibrium,mattes2023entangled,solanki2024exotic,Mukherjee2024correlations,igor_thermodynamics,paulino2024thermodynamicscoupledtimecrystals,PhysRevA.108.023516,paulino2024thermodynamicscoupledtimecrystals,solanki2024chaos, nadolny2024nonreciprocal}. As recently demonstrated, such exotic non-stationary phases can also serve as resources for quantum-enhanced metrology \cite{cabot2023continuous,montenegro2023quantum,gribben2024quantum}. 
For the case of a simple, largely analytically tractable system, our investigation thus highlights how seemingly detrimental incoherent processes may augment coherent dynamics and stabilize collective quantum phenomena.

\paragraph{Model, symmetries and mean-field equations of motion.---} We consider an open BHD model described by the  quantum master equation $\dot \rho = \mathcal{L}[H;\{\mathcal{O}_i\}]\rho$, with  dynamical generator 
\begin{equation}
\mathcal{L}[H;\{\mathcal{O}_i\}]\rho=-i[H,\rho]+\sum_i (\mathcal{O}_i \rho \mathcal{O}_i^\dagger-\frac{1}{2}\{\mathcal{O}_i^\dagger\mathcal{O}_i,\rho\}),
    \label{eq:BH_themal}
\end{equation}
where $H=\Omega(\hat{a}^\dagger \hat{b} + \hat{b}^\dagger \hat{a})/2 + U \sum_{\alpha=a,b}\hat{n}_\alpha (\hat{n}_\alpha-1)$ generates coherent hopping and accounts for interactions.
Here, $\hat{n}_\alpha=\hat{\alpha}^\dagger \hat{\alpha}$ is a number operator with $\alpha \in \{a,b\}$, $\Omega$ represents the coherent hopping rate and $U$ is the on-site interaction strength.
The incoherent hopping processes between modes $a$ and $b$ are implemeneted by the jump operators $\mathcal{O}_1=\sqrt{\gamma_R}\hat{a} \hat{b}^\dagger$ and $\mathcal{O}_2=\sqrt{\gamma_L}\hat{a}^\dagger \hat{b}$. Such processes can be realized by three-wave-mixing with a third bosonic mode, which is connected to a thermal bath (see Appendix \ref{sec:ME} for details).
We therefore parametrize $\gamma_{R,L}$ in terms of the equilibrium occupation number $n_{th}$ of such a third mode, $\gamma_R=(1+n_{th})\kappa$ and $\gamma_L=n_{th}\kappa$, with $\kappa$ setting the overall dissipation rate.
The occupation $n_{th}$ controls the non-reciprocal character of the incoherent hopping, which goes from perfectly unidirectional ($a \rightarrow b$) for $n_{th}=0$ to fully reciprocal  ($a\leftrightarrow b$) in the limit $n_{th} \rightarrow \infty$.
Similar directional and incoherent hopping terms have been recently considered in the context of bosonic transport~\cite{garbe2024}.

The generator (\ref{eq:BH_themal}) exhibits a strong $U(1)$ symmetry  since the total number operator $\hat{N}=\hat{n}_a+\hat{n}_b$ commutes with both the Hamiltonian $H$ and the jump operators $\mathcal{O}_{1,2}$ \cite{Buca2012strong,Albert2014strong}.
Thus, the total number $N=\langle \hat{N} \rangle$ of excitations is conserved.
In what follows, we focus on large excitation numbers, $N\rightarrow\infty$. 
This requires the parameters $U$ and $\kappa$ to be rescaled by $N/2$, such that $U\rightarrow U/(N/2)$ and $\kappa\rightarrow \kappa/(N/2)$, to ensure a consistent thermodynamic limit \cite{kac1963van} and for discussing a spin-equivalent model (see Appendix \ref{sec:BTC}).
The generator in Eq.~(\ref{eq:BH_themal}) also features a strong Lindbladian parity-time ($\mathcal{PT}$) symmetry \cite{nakanishi2024CTCpt}, defined as $\mathcal{L}[\mathcal{PT}(H);\{\mathcal{PT}(\mathcal{O}_i)\}]=\mathcal{L}[H,\{\mathcal{O}_i\}]$,
where $\mathcal{PT}(O):=(PT)O^\dagger (PT)^{-1}$. Here, the parity operator, $\mathcal{P}$, exchanges indices $a \leftrightarrow b$, and the time-reversal operator, $\mathcal{T}$, implements complex conjugation. 
In a non-Hermitian system, $\mathcal{PT}$ symmetry signifies a balance between gain and loss \cite{bender2019pt}.
Recently the notion of $\mathcal{PT}$ symmetry has been extended to Lindblad dynamics~\cite{Prosen_2012,huber2020,nadolny2024nonreciprocal,nakanishi2024CTCpt, nakanishi2022, nakanishi2022dissipative, Roccati_2021}, where it can be broken in the steady state.
Further below, we will show how phases of broken and unbroken $\mathcal{PT}$-symmetry manifest.

\begin{figure*}[htp!]
    \centering
    \includegraphics[width=\linewidth]{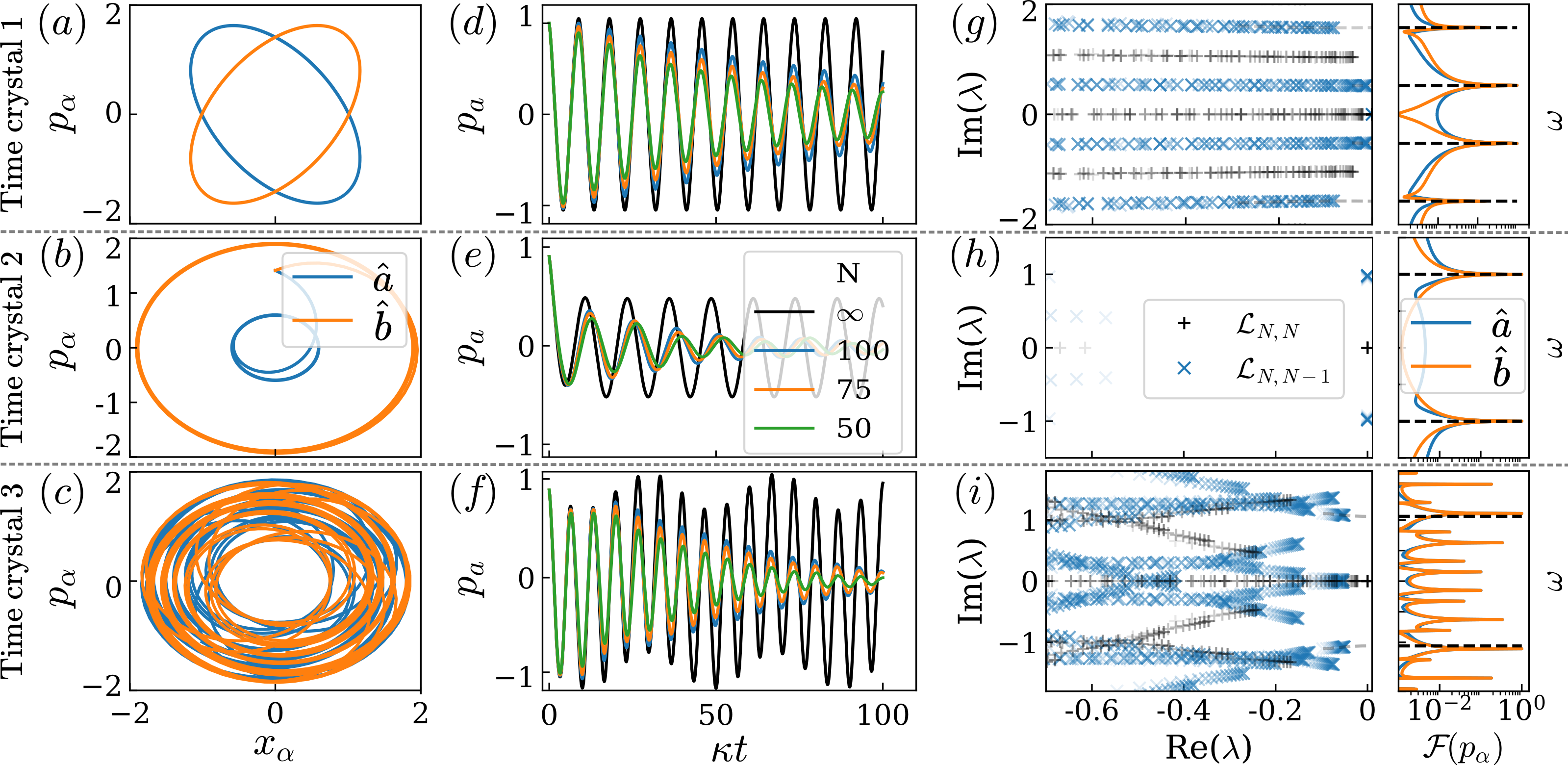}
    \caption{\textbf{Non-stationary phases.} (a-c) Mean-field dynamics of three different time-crystal phases. (d-f) The exact dynamics of Eq.~(\ref{eq:BH_themal}) approaches the corresponding mean-field dynamics (a-c) when increasing the system size. (g-i) Finite-size scaling shows that the dominant eigenvalues of $\mathcal{L}$ (left panel) approach the frequencies of the mean-field variables $p_\alpha$ described by their normalized Fourier transform $\mathcal{F}(p_\alpha)$ (right panel). Color intensities of markers increase with increasing $N$. We consider $N=\{20,25,\ldots, 75\}$. The parameter values for three different time-crystal phases (TC1, TC2, TC3) described by sub-figures (a,d,g), (b,e,h), and (c,f,i) are ($\Omega/\kappa=1.45, U/\kappa=0$), ($\Omega/\kappa=0.8, U/\kappa=0.25$), and ($\Omega/\kappa=1.45, U/\kappa=0.25$), respectively.}
    \label{fig:MF-dynamics}
\end{figure*}

Before getting to this point, we first introduce the mean-field equations that govern the BHD dynamics in the limit $N \rightarrow\infty$:
    \begin{eqnarray}
    \dot{x}_a&=&\Omega p_b/2-\kappa x_a (x_b^2+p_b^2)/4 + U p_a(x_a^2+p_a^2)  ,\nonumber \\
    \dot{p}_a&=&-\Omega x_b/2-\kappa p_a (x_b^2+p_b^2)/4 - U x_a(x_a^2+p_a^2) ,\nonumber \\
    \dot{x}_b&=&\Omega p_a/2+\kappa x_b (x_a^2+p_a^2)/4 + U p_b(x_b^2+p_b^2) ,\nonumber \\
    \dot{p}_b&=&-\Omega x_a/2+\kappa p_b (x_a^2+p_a^2)/4 - U x_b(x_b^2+p_b^2).
\end{eqnarray}
Here $x_\alpha = \langle \hat{x}_\alpha \rangle/\sqrt{N/2} =\langle \hat{\alpha} +\hat{\alpha}^\dagger\rangle/\sqrt{N}$ and $p_\alpha = \langle \hat{p}_\alpha \rangle/\sqrt{N/2} =\langle \hat{\alpha} -\hat{\alpha}^\dagger \rangle/\sqrt{N}i$ with $\alpha\in\{a,b\}$.
Note that we use the hat to distinguish between the quadrature operators and the associated mean-field variables.
The emergence of the mean-field dynamics can be further verified by the time-evolution described by Eq.~(\ref{eq:BH_themal}) for different system sizes, as shown in Fig.~\ref{fig:MF-dynamics}.
Due to the strong $U(1)$ symmetry, the mean-field equations satisfy $\sum_{\alpha=a,b} (x_\alpha^2+p_\alpha^2)/2=2$ and as a result of the $\mathcal{PT}$ symmetry \cite{nakanishi2024CTCpt}, they are invariant under the combined transformation: $a \leftrightarrow b,~~ p_{\alpha}\rightarrow -p_{\alpha},~~ t\rightarrow -t$.
The mean-field equations provide an efficient way to analyze the different phases of the system.

\paragraph{Dissipative phases and phase transitions.---} The system features stationary and non-stationary phases, which are closely linked to the breaking of $\mathcal{PT}$ symmetry. In the absence of the on-site interaction ($U=0$), a stationary phase is reached in the long-time limit, provided that $\Omega/\kappa<1$. When the coherent hopping strength exceeds the incoherent hopping rate, $\Omega/\kappa > 1$, the system enters an oscillatory phase, as illustrated in Fig.~\ref{fig:MF-dynamics}(a).
At the mean-field level, this oscillatory phase is reminiscent of limit cycles in non-linear systems.
In the case $U=0$, our model can be indeed mapped to the well-known boundary time crystal (BTC) through a Schwinger-Boson transformation \cite{schwingertrasnformation1988} (see  Appendix \ref{sec:BTC}).
Therefore, we refer here to the non-stationary phase as the time crystal and the stationary one as a melted phase.
    Figure~\ref{fig:MF-dynamics}(g) shows that sub-systems $a$ and $b$ oscillate with a common frequency.
    Interestingly, the dynamics of the mean-field observable $p_{a,b}$ consists only of odd harmonics, whereas the spin observables of the equivalent BTC model exhibit all the harmonics.
    The time-invariant steady state is not preserved under the $\mathcal{PT}$ symmetry operations,
    whereas the time-crystal phase shown in Fig.~\ref{fig:MF-dynamics}(a) is a $\mathcal{PT}$ symmetric phase.
    Therefore, the system undergoes a $\mathcal{PT}$ phase transition at $\Omega_c/\kappa=1$, and a  $\mathcal{PT}$ symmetric time-crystal phase (named TC1 in Fig.~\ref{fig:MF-dynamics}) emerges for $\Omega/\kappa>1$.

 In the presence of on-site interactions, $U/\kappa  \neq 0$, the system transitions from a limit cycle (TC2) to quasi-periodic dynamics (TC3) at interaction strength $U_c/\kappa=\sqrt{(\Omega/\kappa)^2-1}/4$.
    This behavior is shown in Figs.~\ref{fig:MF-dynamics}(b,c).
    The limit cycles observed for $U>U_c$ are circular in shape and have different radii for sub-systems $a$ and $b$, as shown in Fig.~\ref{fig:MF-dynamics}(b).
    The TC2 oscillations are simple harmonics, and both sub-systems exhibit a single common frequency \cite{supp}, as depicted in Figs.~\ref{fig:MF-dynamics}(e,h).
    The TC2 phase represents a $\mathcal{PT}$ symmetry-broken phase since both sub-systems have different radii, and the system is not invariant under the $\mathcal{PT}$ symmetry operation.
    The radii of the two sub-systems approach the same value at $U_c/\kappa$, and the dynamics becomes quasi-periodic for $U< U_c$, see Fig.~\ref{fig:MF-dynamics}(c).
    The time-averaged radius of both sub-systems approaches the same value for the TC3 phase \cite{supp}, representing the $\mathcal{PT}$ symmetric phase.
    This quasi-periodic behavior is evident in the Fourier transform of the mean-field variables, showing finite incommensurate frequencies that are identical for both the sub-systems, as illustrated in Fig.~\ref{fig:MF-dynamics}(i).
    In summary, the inclusion of a simple incoherent hopping process leads to the emergence of a stationary phase and three distinct time-crystal phases in a BHD.

\begin{figure*}[htp!]
    \centering
    \includegraphics[width=\linewidth]{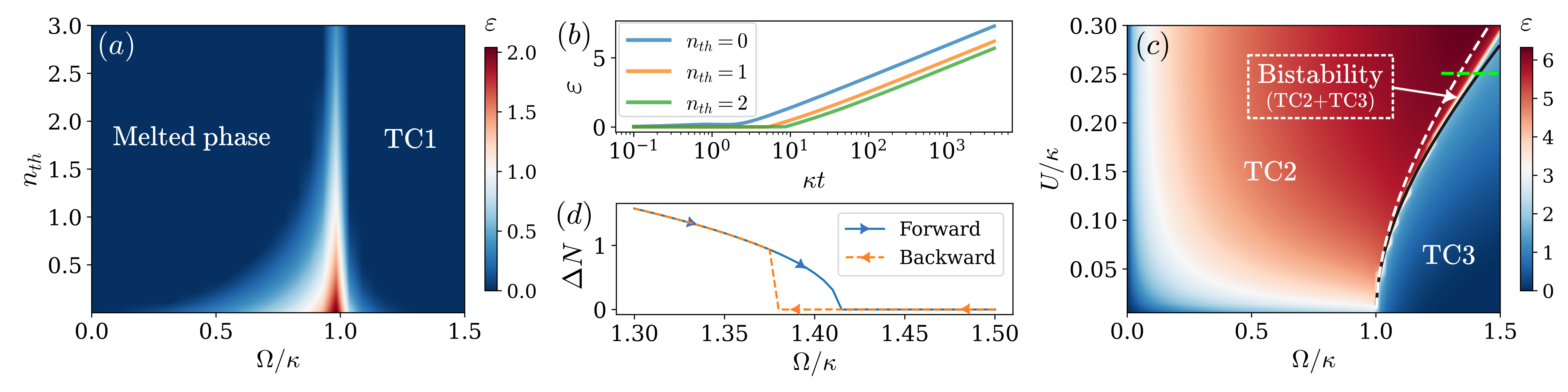}
    \caption{\textbf{Entanglement and bistability.} (a) The time-averaged logarithmic negativity $\varepsilon$ exhibits a maximum at the critical point $\Omega_c/\kappa=1$ for $U/\kappa=0$, and it decreases with increasing $n_{th}$. (b) As time $t$ increases, the entanglement $\varepsilon$ grows at a rate proportional to $\ln(\kappa t)$ when tuned to the critical point $\Omega_c/\kappa$. Increasing $n_{th}$ does not change the slope but merely shifts the onset of entanglement generation. The divergence of correlations indicates a second-order phase transition. (c) For a fixed $U/\kappa \neq 0$, entanglement grows with increasing $\Omega/\kappa$ and decreases rapidly for $\Omega/\kappa$ values beyond the solid (black) line, which represents $U_c/\kappa$ (here $n_{th}=0$). 
    The two time-crystal phases are separated by a bistable region between the dashed (white) and solid (black) lines.
    (d) The order parameter $\Delta N$ displays bistability when adiabatically sweeping $\Omega/\kappa$ in forward and backward directions, keeping $U/\kappa=0.25$ fixed [the parameter regime is indicated by the horizontal (green) dashed line in panel (c)].
    The average entanglement in (a) and (c) is calculated for $\kappa t=4 \times 10^3$.} 
    \label{fig:quantum-correlations}
\end{figure*}
\paragraph{Liouvillian eigenspectrum analysis.---} We now investigate how the mean-field dynamics emerges from the (linear) Lindblad time evolution in the limit $N\rightarrow \infty$.
In any experimental realization of a BHD, the number of excitations is finite.
Consequently, we investigate the effect of increasing system size on the dynamics of the system.
We exploit the $U(1)$ symmetry and the conservation of the total number of excitations to block-diagonalize $\mathcal{L}$ in Eq.~\eqref{eq:BH_themal} and calculate its spectrum. 
We define the basis $\vert N,N_a \rangle$ where $N_{a,b}$ is the total number of excitations for sub-systems $a$ and $b$  such that $N=N_a+N_b$  is conserved.
The density matrix can then be expressed as
\begin{equation}
\rho(t)=\sum_{N,N'}\sum_{N_a=0}^N\sum_{N'_a=0}^{N'}\rho^{N,N'}_{N_a,N'_a}(t)|N,N_a\rangle\langle N',N'_a|,
\end{equation}
where $N$ and $N^{'}$ label two different excitation subspaces.
The Liouvillian $\mathcal{L}$ displays a block-diagonal structure $\mathcal{L}=\bigoplus_{N,N'}\mathcal{L}_{N,N'}$, where each possible pair $\{N,N'\}$ defines an independent block.
The expectation value of creation and annihilation operators solely depends on the density matrix elements between neighboring sectors $\{N,N\pm1\}$ (see  Appendix \ref{sec:spectra}). Thus, the emergence of persistent oscillations in quantities like $\hat{x}_{a,b}$ and $\hat{p}_{a,b}$ must be signaled by some eigenvalues of $\mathcal{L}_{N,N\pm1}$ becoming purely imaginary in the thermodynamic limit (see Appendix \ref{sec:spectra}).
We investigate the eigenspectra of these blocks in relation to the time-crystal phases.

Let us first discuss the TC1 phase ($U=0$ and $\Omega/\kappa>1)$. 
Figure~\ref{fig:MF-dynamics}(g) shows that the spectral gap closes upon increasing the system size.
Interestingly, only the eigenvalues of  $\mathcal{L}_{N,N-1}$ corresponding to the odd harmonics present in the Fourier transform of $p_\alpha$ become gapless in agreement with the mean-field behavior.  

In the parameter regime $U>U_c$, the spectral gap of $\mathcal{L}_{N,N^{'}}$ closes only for subspaces with $N^{'}=N- 1$. 
This leads to the TC2 time-crystal phase with a single purely imaginary eigenvalue that matches the frequency of the mean-field oscillations, as shown in Fig.~\ref{fig:MF-dynamics}(h).
In the opposite regime ($U<U_c$), we observe a quasi-periodic time-crystal phase with various frequencies.
In Fig.~\ref{fig:MF-dynamics}(i), we show that the imaginary part of the first dominant eigenvalues of $\mathcal{L}_{N,N-1}$ approaches one of the prominent frequencies of the mean-field oscillations. 
Due to finite-size effects, it is challenging to isolate all the eigenvalues that become purely imaginary in the thermodynamic limit.
The real part of such eigenvalues might vanish at different rates with system size \cite{cabot2023nonequilibrium,dutta2024quantumoriginlimitcycles,solanki2024chaos}, making it challenging to observe those frequencies. 
Finite-size analysis of the Liouvillian eigenspectrum thus shows how the mean-field phases emerge and remain stable as the system size diverges.

\paragraph{Quantum correlations.---} 
After establishing the existence of different nonequilibrium phases, we now take a step further and explore quantum correlations to further elucidate the nature of phase transitions. 
In particular, we investigate whether and how the presence of stochastic incoherent hopping can impact the generation of entanglement.
Investigating quantum correlations also allows us to determine the nature of the different phase transitions observed.
To capture the quantum correlations between the sub-systems, we consider the behavior of the fluctuations around the mean-field solutions \cite{boneberg2022quantum,mattes2023entangled}. 
Fluctuation operators are defined as $F^{\alpha}=T^\alpha-\langle T^\alpha \rangle$, where $T^\alpha$ denotes the $\alpha$-th component of the tuple $T=(\hat{x}_a,\hat{p}_a,\hat{x}_b,\hat{p}_b)$. 
In the thermodynamic limit, quantum fluctuations are Gaussian and hence fully described by the covariance matrix $\Sigma^{\alpha \beta} = \langle \{F^{\alpha},F^{\beta}\} \rangle /2$ \cite{boneberg2022quantum,mattes2023entangled}. This allows us to characterize correlations using the quantum information toolbox for Gaussian states \cite{olivares2012quantum,adesso2014continuous}. The dynamics of the covariance matrix is governed by the following Lyapunov (matrix) equation,
\begin{equation}\label{eq:Lyapunov}
\dot{\Sigma}(t)= \Sigma(t)(A^T+Q^T)+(A+Q)\Sigma(t) + (Z+Z^T)/2,
\end{equation}
where the expression for the matrices $A$, $Q$, and $Z$ can be found in Appendix \ref{sec:fluctuations}. We note that these matrices can be time-dependent, for instance, when considering the fluctuations around a time-crystal phase \cite{carollo2022exact}.
The Lyapunov equation (\ref{eq:Lyapunov}) is derived in Appendix \ref{sec:fluctuations} by computing the equations of motion for the fluctuation operators using the master equation and keeping only the dominant terms in the thermodynamic limit \cite{boneberg2022quantum,mattes2023entangled}.

We use the logarithmic negativity \cite{adesso2005entanglement,mattes2023entangled} to quantify the entanglement properties as a function of the parameters $\Omega/\kappa$, $U/\kappa$, and $n_{th}$ (see Fig.~\ref{fig:quantum-correlations}).
First, we consider $U=0$. 
The entanglement between the two modes diverges logarithmically in time at the phase transition point $\Omega_c/\kappa$ as shown in Fig.~\ref{fig:quantum-correlations}(a).
The dynamics of the system remain independent of the asymmetry in the incoherent processes, i.e., on $n_{th}$, but the amount of entanglement decreases with increasing $n_{th}$.
Interestingly, the entanglement between modes $a$ and $b$ diverges even at finite $n_{th}$ with the same logarithmic scaling as at $n_{th}=0$, see Fig.~\ref{fig:quantum-correlations}(b).
The effect of finite $n_{th}$ is reflected in the delay of the onset of the entanglement between two modes.
The critical behavior can be further understood by looking at the scaling of the order parameter $\Delta N= \vert \langle \hat{a}^\dagger \hat{a} -\hat{b}^\dagger \hat{b} \rangle\vert/2$ near the phase transition point. Here, we find that $\Delta N \propto [(\Omega_c-\Omega)/\kappa]^{1/2}$ (see Appendix \ref{sec:entanglement}), revealing a second-order phase transition with a static critical exponent of $1/2$. Further evidence for such a transition is given by the divergence of correlations.

For $U\neq 0$, the system features bistability between the time-crystal phases TC2 and TC3 (see Appendix \ref{sec:bistable}), which suggests that the second-order transition becomes a first-order one.
The bistability exists within the region bounded by the dashed (white) and the solid (black) line (defined by $U_c/\kappa$) in Fig.~\ref{fig:quantum-correlations}(c).
The solid line represents the transition from the TC2 phase to the TC3 phase, achieved by starting in the TC2 phase at a lower value of $\Omega/\kappa$ and then gradually increasing its value while keeping $U/\kappa$ fixed.
Conversely, the transition from the TC3 to the TC2 phase occurs at distinct values of parameters, indicated by the white dashed line in Fig.~\ref{fig:quantum-correlations}(c), which is obtained by starting with the TC3 phase and then adiabatically decreasing the $\Omega/\kappa$ values.
This bistability gives rise to the hysteresis behavior of the order parameter $\Delta N$, which is depicted in Fig.~\ref{fig:quantum-correlations}(d) for a fixed value of $U/\kappa=0.25$.
The transition between these two different time-crystal phases at $U_c/\kappa$ is captured by the change in the entanglement between the two modes.
This indicates that the single-frequency time-crystal phase, which emerges when $U>U_c$, exhibits large entanglement, in contrast to the quasi-periodic dynamics observed when  $U<U_c$.
Note that $n_{th}=0$ for Fig.~\ref{fig:quantum-correlations}(c).
Increasing $n_{th}$ reduces the maximally achievable entanglement while the qualitative features of the phase diagram remain unchanged.
The saturation of entanglement in time at $U_c/\kappa$, along with the existence of bistability, indicates a first-order phase transition.

\paragraph{Conclusion.---} Using the BHD as a case study, we have shown how tunable parameters like interactions, coherent hopping, and incoherent hopping rates enable precise control over quantum correlations, such as entanglement, which is a key resource for quantum metrology. 
Remarkably, we find that incorporating simple incoherent hopping processes not only enhances entanglement generation but also gives rise to exotic non-stationary phases, including time-crystal behavior, which have found their application in quantum sensing \cite{montenegro2023quantum,cabot2023continuous} and quantum thermodynamics \cite{igor_thermodynamics,paulino2024thermodynamicscoupledtimecrystals}.
Our findings thus contribute to the broader quest to harness and control emergent collective behavior via simple, coherent, and incoherent processes in quantum many-body systems.

\paragraph{Acknowledgments.---}  P.S. acknowledges support from the Alexander von Humboldt Foundation through a Humboldt research fellowship for postdoctoral researchers. C.B. and P.S. acknowledge financial support from the Swiss National Science Foundation individual grant (grant no. 200020 200481). F.C.~is indebted to the Baden-W\"urttemberg Stiftung for the financial support of this research project by the Eliteprogramme for Postdocs. A.C. acknowledges support from the Deutsche Forschungsgemeinschaft (DFG, German Research Foundation) through the Walter Benjamin programme, Grant No. 519847240. 
M.B. acknowledges funding from
the European Research Council (ERC) under the European Union’s Horizon 2020 research and innovation program (Grant agreement No. 101002955 - CONQUER). This work was supported by the QuantERA II programme (project CoQuaDis, DFG Grant No. 532763411) that has received funding from the EU H2020 research and innovation programme under GA No. 101017733.  We also acknowledge funding from the Deutsche Forschungsgemeinschaft (DFG, German Research Foundation) through the Research Unit FOR 5413/1, Grant No.~465199066.

\appendix
\onecolumngrid
\renewcommand{\thefigure}{S\arabic{figure}}
\setcounter{figure}{0} 

\section{Derivation of the master equation \label{sec:ME}}

Here, we give a detailed derivation of Eq.~(\ref{eq:BH_themal}) of the main text.
We assume that modes $a$ and $b$ are interacting with a third mode $c$, which is in contact with a thermal bath. The system is described by the master equation
\begin{eqnarray}
    \dot \rho= -i [H_0+H_{ab}(t) + \Tilde{g} (\hat{a}^\dagger \hat{b} \hat{c}+\hat{a}\hat{b}^\dagger \hat{c}^\dagger), \rho]+\gamma (1+n_{th})\mathcal{D}[\hat{c}] \rho+\gamma n_{th}\mathcal{D}[\hat{c}^\dagger ] \rho.
    \label{eq:abc_me_time}
\end{eqnarray}
The bare Hamiltonian is given by $H_0=\sum_{\alpha=a,b,c}\omega_\alpha \hat{n}_\alpha + (2U/N) \sum_{\alpha=a,b}\hat{n}_\alpha (\hat{n}_\alpha-1)$, where $\omega_\alpha$ represents the frequency of each mode, with the condition $\omega_c=\omega_a-\omega_b$ for the three-wave mixing process.
The interaction between modes $a$ and $b$ is described by the parametric coupling term $H_{ab}(t)=\Omega( \hat{a} \hat{b}^\dagger e^{i\omega_d t}+\hat{a}^\dagger \hat{b} e^{-i\omega_d t})/2 $, where $\omega_d=\omega_a-\omega_b$ ensures coherent exchange between modes $a$ and $b$. 
In the three-wave mixing term, the coupling constant scales as $\Tilde{g}=g/\sqrt{N/2}$ to guarantee the correct thermodynamic scaling. 
In the rotated frame defined by $\mathcal{U}=\exp(i\sum_{\alpha=a,b,c}\omega_\alpha \hat{n}_\alpha t)$, Eq.~(\ref{eq:abc_me_time}) becomes time-independent and reads
\begin{eqnarray}
    \dot \rho=(\mathcal{L}_0+\mathcal{L}_1) \rho= \underbrace{{-i [H + \Tilde{g} (\hat{a}^\dagger \hat{b} \hat{c}+\hat{a}\hat{b}^\dagger \hat{c}^\dagger), \rho]}}_{\mathcal{L}_1\rho}+\underbrace{\gamma (1+n_{th})\mathcal{D}[\hat{c}] \rho+\gamma n_{th}\mathcal{D}[\hat{c}^\dagger ] \rho}_{\mathcal{L}_0 \rho}.
    \label{eq:abc_me}
\end{eqnarray}
where $H=\Omega( \hat{a} \hat{b}^\dagger +\hat{a}^\dagger \hat{b})/2 + (2U/N) \sum_{\alpha=a,b}\hat{n}_\alpha (\hat{n}_\alpha-1)$.

In the strong dissipation limit $\gamma\gg \Omega,U,g$, the dynamics is fast under $\mathcal{L}_0$ and slow for $\mathcal{L}_1$. 
Therefore, we can adiabatically eliminate \cite{breuer2002theory,gonzalez2024tutorial,krishna2023select,kinetic2013igor} the mode $c$ to formulate an effective master equation for the dynamics of the sub-system formed by $a$ and $b$.

Let us first discuss the system's time evolution under the fast dynamics. 
Since $\mathcal{L}_0$ models a thermal bath, the mode $c$ will settle down to a thermal state in the long-time limit, 
\begin{equation}
    \mathcal{P}\rho=\lim_{t\rightarrow\infty} e^{\mathcal{L}_0 t} \rho=Tr_c[\rho]\otimes \rho_{th}.
\end{equation}
Here $\mathcal{P}$ is the projection operator to the steady state of the thermal map.

Now, the effective master equation for the slow dynamics of the projected subspace $\mu =\mathcal{P}\rho$ is given by
\begin{equation}
    \Dot{\mu}= \mathcal{P} \mathcal{L}_1 \mu + \int_0^\infty dt \mathcal{P}\mathcal{L}_1\mathcal{Q} e^{\mathcal{L}_0 t}  \mathcal{Q} \mathcal{L}_1 \mu = \mathcal{L}_{\text{eff}} \mu 
\end{equation}
where $\mathcal{Q}=I-\mathcal{P}$ projects on the complement subspace. 
The above equation can be further simplified using $\mathcal{Q} \mathcal{L}_1 \mathcal{P} =\mathcal{L}_1 \mathcal{P} $. 
The dynamics of sub-systems $a$ and $b$ after tracing out mode $c$ is given by
\begin{equation}
    \Dot{\mu}_s=Tr_c[\mathcal{P} \mathcal{L}_1 \mu + \int_0^\infty dt \mathcal{P}\mathcal{L}_1 e^{\mathcal{L}_0 t}  \mathcal{L}_1 \mu] = \mathcal{L}_{\text{eff}} \mu_s , \label{eq:ab_ME}
\end{equation}
where $\mu_s=Tr_c[\mu]$. Now, let us calculate $\mathcal{P}\mathcal{L}_1 e^{\mathcal{L}_0 t}  \mathcal{L}_1 \mu$, which can be simplified as follows,

\begin{align}
  \mathcal{P}\mathcal{L}_1 e^{\mathcal{L}_0 t}  \mathcal{L}_1 \mu = \Tilde{g}^2\mathcal{P}Tr_c&[ -\hat{a}^\dagger \hat{b} \hat{a} \hat{b}^\dagger \{\hat{c} e^{\mathcal{L}_0 t} (\hat{c}^\dagger \mu)\} + \hat{a} \hat{b}^\dagger \{ e^{\mathcal{L}_0 t} (\hat{c}^\dagger \mu) \hat{c}\}  \hat{a}^\dagger \hat{b}   -\hat{a} \hat{b}^\dagger \hat{a}^\dagger \hat{b} \{\hat{c}^\dagger e^{\mathcal{L}_0 t} (\hat{c} \mu)\} + \hat{a}^\dagger \hat{b} \{ e^{\mathcal{L}_0 t} (\hat{c} \mu) \hat{c}^\dagger \}  \hat{a} \hat{b}^\dagger \nonumber \\ 
  &+\hat{a}^\dagger \hat{b} \{\hat{c} e^{\mathcal{L}_0 t} (\mu \hat{c}^\dagger )\} \hat{a} \hat{b}^\dagger  - \{ e^{\mathcal{L}_0 t} (\mu \hat{c}^\dagger ) \hat{c}\}  \hat{a} \hat{b}^\dagger  \hat{a}^\dagger \hat{b}   +\hat{a} \hat{b}^\dagger \{\hat{c}^\dagger e^{\mathcal{L}_0 t} (\mu \hat{c} )\} \hat{a}^\dagger \hat{b}  - \{ e^{\mathcal{L}_0 t} (\mu \hat{c} ) \hat{c}^\dagger \} \hat{a}^\dagger \hat{b}   \hat{a} \hat{b}^\dagger]  .\label{eq:adibatic_c}
\end{align}
Here, each term inside $\{.\}$ can be calculated using the two-time correlation functions 
\begin{align}
    \langle O_1 (0) O_2(t)\rangle = Tr_c[O_2 (0) e^{\mathcal{L}_0 t}(\mu (t) O_1(0)) ], ~~~ \langle O_1 (t) O_2(0)\rangle = Tr_c[O_1 (0) e^{\mathcal{L}_0 t}(O_2(0)\mu (t)) ].
\end{align}
Using these correlation functions and the cyclic property of the trace,  Eq.~(\ref{eq:adibatic_c}) can be rewritten as 
\begin{align}
  \mathcal{P}\mathcal{L}_1 e^{\mathcal{L}_0 t}  \mathcal{L}_1 \mu = \Tilde{g}^2 &[ -\hat{a}^\dagger \hat{b} \hat{a} \hat{b}^\dagger \mu_s \langle \hat{c}(t) \hat{c}^\dagger (0) \rangle + \hat{a} \hat{b}^\dagger \mu_s \hat{a}^\dagger \hat{b} \langle \hat{c}(t) \hat{c}^\dagger (0) \rangle    -\hat{a} \hat{b}^\dagger \hat{a}^\dagger \hat{b} \mu_s \langle  \hat{c}^\dagger (t) \hat{c}(0) \rangle + \hat{a}^\dagger \hat{b} \mu_s  \hat{a} \hat{b}^\dagger \langle  \hat{c}^\dagger (t) \hat{c}(0) \rangle   \nonumber \\ 
  &+ \hat{a}^\dagger \hat{b} \mu_s \hat{a} \hat{b}^\dagger \langle  \hat{c}^\dagger (0) \hat{c}(t) \rangle -\mu_s \hat{a} \hat{b}^\dagger  \hat{a}^\dagger \hat{b} \langle  \hat{c}^\dagger (0) \hat{c}(t) \rangle   +\hat{a} \hat{b}^\dagger \mu_s \hat{a}^\dagger \hat{b} \langle  \hat{c}(0) \hat{c}^\dagger (t) \rangle - \mu_s \hat{a}^\dagger \hat{b} \hat{a} \hat{b}^\dagger  \langle \hat{c}(0) \hat{c}^\dagger (t) \rangle]  .
\end{align}

The time evolution of operator $\hat{c}$ and $\hat{c}^\dagger$ is defined as follows
\begin{equation}
    \partial_t \hat{c} =\mathcal{L}_0^* \hat{c}= -\gamma \hat{c}/2, ~~~ ~~~~ \partial_t c^\dagger  =\mathcal{L}_0^* \hat{c}^\dagger = -\gamma \hat{c}^\dagger /2.
\end{equation}
The above equations have solutions of the form $\hat{c}(t)=e^{-\gamma t/2}\hat{c}(0)$ and $\hat{c}^\dagger(t)=e^{-\gamma t/2}\hat{c}^\dagger(0)$. 
Substituting these solutions along with Eq.~(\ref{eq:adibatic_c}) in Eq.~(\ref{eq:ab_ME}), the reduced master equation for $\mu_s$ becomes
\begin{equation}
    \Dot{\mu}_s = -i[H,\mu_s]+ \frac{2\kappa}{N}(1+n_{th})\mathcal{D}[\hat{a}\hat{b}^\dagger ] \mu_s + \frac{2\kappa}{N}n_{th}\mathcal{D}[\hat{a}^\dagger \hat{b} ]\mu_s, \label{eq:ME_ab}
\end{equation}
where $\kappa=4g^2/\gamma$.

\section{Mean-field equations and fixed points \label{sec:MF}}

To get analytical insights for the mean-field dynamics, we investigate the time evolution of the complex variables $\alpha= \langle \hat{a} \rangle/\sqrt{N/2} = \langle \hat{x}_a+i\hat{p}_a \rangle/\sqrt{N} $ and $\beta =\langle \hat{b} \rangle/\sqrt{N/2}=\langle \hat{x}_b+i\hat{p}_b \rangle/\sqrt{N} $.
The mean-field dynamics of these variables is described by the following coupled equations
\begin{eqnarray}
\dot{\alpha}=-i\frac{\Omega}{2}\beta-\frac{\kappa}{2}\alpha |\beta|^2-i2U\alpha|\alpha|^2,\\
\dot{\beta}=-i\frac{\Omega}{2}\alpha+\frac{\kappa}{2}\beta |\alpha|^2-i2U\beta|\beta|^2.
\end{eqnarray}
The above mean-field equations can also be expressed in terms of the quadratures $x_a=\sqrt{2}\text{Re}[\alpha]$, $p_a=\sqrt{2}\text{Im}[\alpha]$, $x_b=\sqrt{2}\text{Re}[\beta]$ and $p_b=\sqrt{2}\text{Im}[\beta]$, which have been discussed in the main text.

We can further decompose the dynamics of the complex variables $\alpha$ and $\beta$ using polar coordinates
\begin{equation}
\alpha=R_a e^{i\phi_a},\quad \beta=R_b e^{i\phi_b}, \quad \Delta \phi=\phi_b-\phi_a,\quad \Sigma_\phi=\phi_a+\phi_b,
\end{equation}
where $R_{a(b)}$ and $\phi_{a(b)}$ represent the magnitude and phase for variable $\alpha$($\beta$), $\Delta \phi$ defines the relative phase and $\Sigma_\phi$ is the total phase.

The dynamics of the system in terms of these polar coordinates is described by the following equations,
\begin{align}
\partial_t{R}_a&=\frac{\Omega}{2} R_b \sin\Delta \phi-\frac{\kappa}{2}R_aR_b^2,\\
\partial_t{R}_b&=-\frac{\Omega}{2} R_a \sin\Delta \phi+\frac{\kappa}{2}R_bR_a^2,\\
\partial_t{\Delta \phi}&=\bigg[\frac{\Omega}{2R_aR_b}\cos\Delta\phi-2U \bigg](R_b^2-R_a^2),\\
\partial_t \Sigma_\phi&=-\bigg[\frac{\Omega}{2R_aR_b}\cos\Delta\phi+2U \bigg](R_b^2+R_a^2).
\end{align}
Interestingly, $R_a$, $R_b$ and $\Delta \phi$ are decoupled from $\Sigma_\phi$.

{\bf Fixed-point analysis:} 
We now obtain the fixed-point solutions for $R_a$, $R_b$ and $\Delta \phi$. 
When $\delta_t \Sigma_\phi=0$, the system settles down to a steady state, while $\delta_t \Sigma_\phi\neq0$ implies synchronized limit cycles.
Moreover, we recall that $|\alpha|^2+|\beta|^2=R_a^2 +R_b^2$ is conserved by the dynamics. In the following, we consider the solutions for the maximum value of this quantity, i.e. $R_a^2 +R_b^2=2$.
The possible fixed-point solutions for $\partial_t{R}_a=\partial_t{R}_b=\partial_t{\Delta \phi}=0$ are 
\begin{equation}
\Delta\phi=\tan^{-1}\bigg( \frac{\kappa}{4U}\bigg),\quad 0\leq \Delta \phi \leq\frac{\pi}{2},
\end{equation}
\begin{equation}
R_a=\sqrt{1\pm\sqrt{1-\frac{\Omega^2}{16U^2+\kappa^2}}},\quad R_b=\sqrt{1\mp\sqrt{1-\frac{\Omega^2}{16U^2+\kappa^2}}}.
\label{eq:mf_solutions}
\end{equation}

These solutions exist only when
\begin{equation}
\Omega^2\leq 16U^2+\kappa^2,
\end{equation}
such that $R_a$ and $R_b$ approach the same value in the limit $\Omega^2 \rightarrow 16U^2+\kappa^2$.
For $U\neq0$ they correspond to limit cycles as the solution for the total phase is
\begin{equation}
\Sigma_\phi(t)=-8Ut+\Sigma_\phi(0).
\end{equation}
Moreover, from this we can obtain the time evolution of $\phi_{a,b}$:
\begin{eqnarray}
\phi_a(t)=-4Ut-\frac{1}{2}\tan^{-1}\bigg( \frac{\kappa}{4U}\bigg)+\phi_a(0),\\
\phi_b(t)=-4Ut+\frac{1}{2}\tan^{-1}\bigg( \frac{\kappa}{4U}\bigg)+\phi_b(0).
\end{eqnarray}
These last two equations tell us that the limit cycles contain just one frequency without higher harmonics. The frequency is $4U$. Finally, we note that when $\Omega^2>16U^2+\kappa^2$, these fixed points cease to be valid solutions, and the system displays quasi-periodic oscillations.

\section{Bistability in the presence of on-site interaction \label{sec:bistable}}

We now discuss the existence of bistability in our model when $U\neq 0$. 
Based on the fixed-point analysis presented in the previous section, the system exhibits simple harmonic oscillations when $U>U_c$ where $U_c/\kappa=\sqrt{(\Omega/\kappa)^2-1}/4$.
In this case, the sub-systems $a$ and $b$ oscillate with distinct amplitudes, $R_a$ and $R_b$, which converge to the same value $R_a=R_b=1$ at $U_c/\kappa$.
The dynamics becomes quasi-periodic for  $U<U_c$, as shown in Figs.~\ref{fig:MF-dynamics}(c,f) of the main text.
The time-averaged amplitude of oscillations for sub-systems $a$ and $b$ is the same for the quasi-periodic phase, i.e, $\bar{R}_a=\bar{R}_b=1$ where $\bar{R}_{a,b} = \frac{1}{T}\int_0^T dt R_{a,b} $ represents the time averaged amplitude.
Thus, the difference between the time-averaged amplitudes of the sub-systems, $\Delta \bar{R}=\vert \bar{R}_a - \bar{R}_b \vert/\sqrt{2}$, acts as an order parameter.
It signals the transition from a single-frequency limit cycle, where $\Delta \bar{R} \neq 0$, to quasi-periodic dynamics, where $\Delta \bar{R}=0$. 

\begin{figure}[b!]
    \centering
    \includegraphics[width=0.8\textwidth]{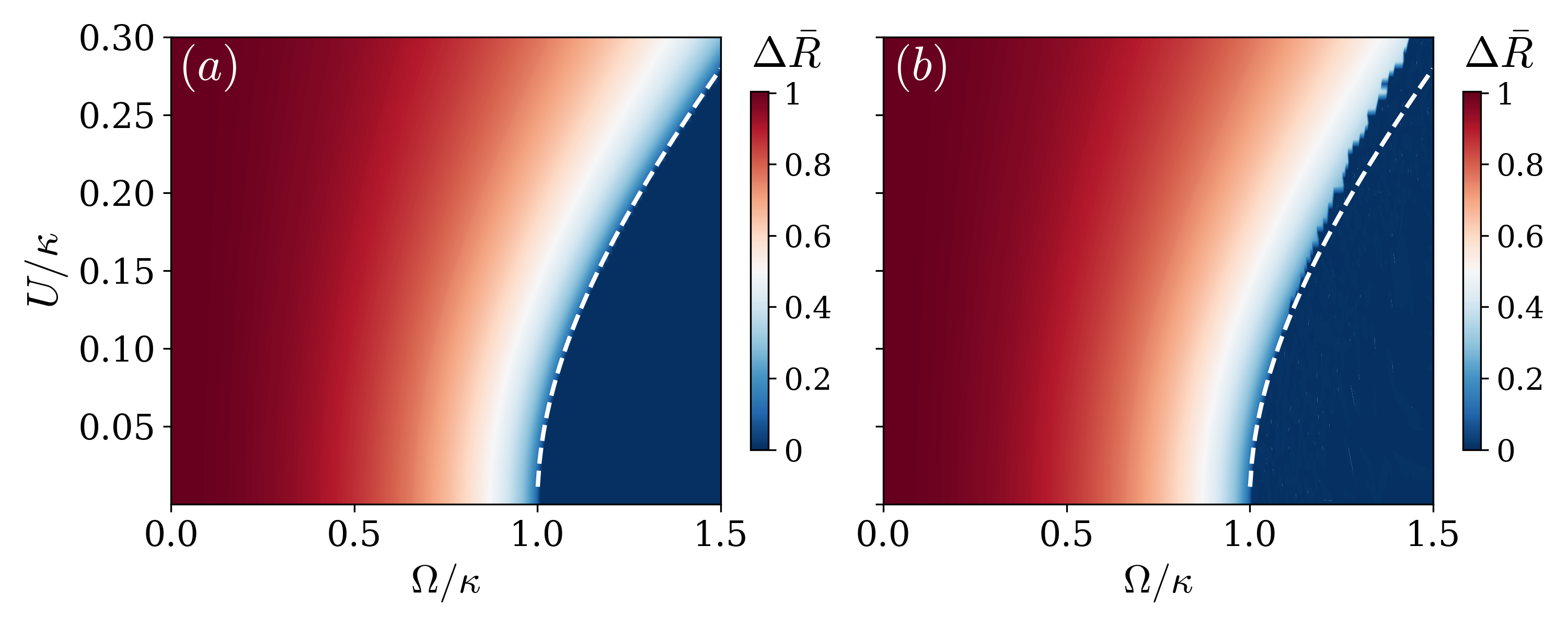}
    \caption{(a) Order parameter $\Delta \bar{R} = |\bar{R}_a - \bar{R}_b|/\sqrt{2}$, showing the phase transition from TC2 ($U>U_c$) to TC3 ($U<U_c$) as $\Omega/\kappa$ increases for fixed $U/\kappa$. The transition occurs at $U_c/\kappa$ (dashed white line) obtained from the fixed-point analysis. (b) The transition from TC3 to TC2 occurs as $\Omega/\kappa$ decreases, with higher $U/\kappa$ values exhibiting bistability beyond the dashed line.}
    \label{fig:bistability}
\end{figure}

In our model, we observe that even in the parameter regime where a single-frequency limit cycle is stable, the system may transit to quasi-periodic dynamics.
Previous observations of bistability involved either two stationary fixed points \cite{PhysRevLett.111.113901,PhysRevLett.113.210401} or a combination of a stationary fixed point and a time-crystal phase \cite{mattes2023entangled}. 
In contrast, the bistability in our model arises from the coexistence of two distinct time-crystal phases.
To illustrate this bistability, we time-evolve the system with a constant value of $U/\kappa$ and then adiabatically sweep the values of $\Omega/\kappa$. 
Depending on the direction of the sweep, the system transits between two different time crystal phases at different values of $\Omega/\kappa$, as shown in Fig.~\ref{fig:bistability}.

For example, we first fix the value of $U/\kappa$ and start with $\Omega/\kappa \ll \sqrt{16 (U/\kappa)^2+1}$. 
Then, we let the system evolve into the corresponding time-crystal phase TC2 and then increase the value of $\Omega/\kappa$. 
Following this adiabatic increase of the parameter $\Omega/\kappa$, the system will transit to quasi-periodic dynamics at $U_c/\kappa$.
This behavior is shown in Fig.~\ref{fig:bistability}(a), where the order parameter $\Delta \bar{R}$ becomes zero only at $U_c/\kappa$.

Now, we sweep the parameter $\Omega/\kappa$ in the reverse direction, keeping $U/\kappa$ fixed.
We begin with $\Omega/\kappa \gg \sqrt{16 (U/\kappa)^2+1}$ and let the system settle down to the TC3 phase with quasi-periodic dynamics.
Now, we adiabatically decrease the value of $\Omega/\kappa$.
Figure~\ref{fig:bistability}(b) shows that for sufficiently large values of $U/\kappa$, the system transits to the time-crystal phase TC2 at a value of $U/\kappa$ that differs from $U_c/\kappa$.
This confirms the coexistence of two time-crystal phases in our model.

\section{Spectra of the Bosonic Liouvillian \label{sec:spectra}}

Here, we discuss the block diagonal structure of $\mathcal{L}$ using the $U(1)$ symmetry of the system.
Since the total excitation $N=N_a+N_b$ of the system is conserved throughout the dynamics,  $\mathcal{L}$ features a block-diagonal structure with respect to these excitation sub-spaces.
In terms of the Fock basis $|N,N_a \rangle$ ($N_b$ is omitted since $N$ is conserved), the state of the system can always be written as
\begin{equation}
\rho(t)=\sum_{N,N'}\sum_{N_a=0}^N\sum_{N'_a=0}^{N'}\rho^{N,N'}_{N_a,N'_a}(t)|N,N_a\rangle\langle N',N'_a|, \,\,\, \text{with} \,\,\, N_b=N-N_a,\,\,\, N'_b=N'-N'_a.
\end{equation}
In terms of this parametrization, the master equation reads:
\small
\begin{equation}
\begin{split}
\partial_t \rho^{N,N'}_{N_a,N'_a}=&-i\frac{\Omega}{2}\bigg[\sqrt{N_a(N_b+1)}\rho^{N,N'}_{N_a-1,N'_a}+\sqrt{N_b(N_a+1)}\rho^{N,N'}_{N_a+1,N'_a}-\sqrt{N'_b(N'_a+1)}\rho^{N,N'}_{N_a,N'_a+1}-\sqrt{N'_a(N'_b+1)}\rho^{N,N'}_{N_a,N'_a-1} \bigg]\\
&-i\frac{2U}{N}\bigg[N_a(N_a+1)+N_b(N_b+1)-N'_a(N'_a+1)-N'_b(N'_b+1) \bigg]\rho^{N,N'}_{N_a,N'_a}\\
&+\frac{2\kappa}{N}(1+n_{th})\bigg[\sqrt{(N_a+1)N_b(N'_a+1)N'_b}\,\rho^{N,N'}_{N_a+1,N'_a+1}-\frac{1}{2} \big(N_a(N_b+1)+N'_a(N'_b+1)\big) \rho^{N,N'}_{N_a,N'_a}\bigg]\\
&+\frac{2\kappa}{N}n_{th}\bigg[\sqrt{(N_b+1)N_a(N'_b+1)N'_a}\,\rho^{N,N'}_{N_a-1,N'_a-1}-\frac{1}{2} \big(N_b(N_a+1)+N'_b(N'_a+1)\big) \rho^{N,N'}_{N_a,N'_a}\bigg].
\end{split}
\end{equation}
\normalsize

From the above equation, it is clear that the Liouvillian is block diagonal and each possible pair of values $\{N,N'\}$ defines an independent block, such that,
\begin{equation}
\mathcal{L}=\bigoplus_{N,N'}\mathcal{L}_{N,N'}.
\end{equation}
Each of the blocks can be diagonalized independently. For finite $N$, only the spectrum of the blocks $\mathcal{L}_{N,N}$ contains a zero eigenvalue. Moreover, the eigenvalues of the $\mathcal{L}_{N,N}$ block coincide with the ones of the corresponding collective spin master equation [Eq. (\ref{eq:BTC_themal})]  for a single total angular momentum sector ($S=N/2$), see below.

We now consider the expected value of the creation and annihilation operators:
\begin{equation}
\text{Tr}[a^\dagger\rho(t)]=\sum_N \sum_{N_a=0}^{N} \sqrt{N_a+1}\, \rho_{N_a,N_a+1}^{N,N+1}(t),\quad\text{Tr}[a\rho(t)]=\sum_N \sum_{N_a=0}^{N} \sqrt{N_a}\, \rho_{N_a,N_a-1}^{N,N-1}(t). 
\end{equation}
They only depend on matrix elements between neighboring sectors $\{N,N\pm1\}$. Thus, the emergence of persistent oscillations  in quantities like $x_{a,b}$, $p_{a,b}$ should be signaled by some eigenvalues of $\mathcal{L}_{N,N\pm1}$ becoming purely imaginary in the thermodynamic limit. When considering the corresponding collective spin model [Eq. (\ref{eq:BTC_themal})], this means that the limit cycles observed in $x_{a,b}$, $p_{a,b}$ do not manifest in a single total angular momentum sector, but require initial states that contain at least two neighboring total angular momentum sectors $S$ and $S\pm1/2$.

\section{Dynamics of quantum fluctuation operators \label{sec:fluctuations}} 

To investigate the nature of correlations between modes $a$ and $b$, we investigate the dynamics of quantum fluctuation operators \cite{boneberg2022quantum,mattes2023entangled}, defined as follows
\begin{eqnarray}
 F^{\alpha} = \hat{\alpha} - \langle \hat{\alpha} \rangle,
\end{eqnarray}
where $\alpha=x_{a},p_{a},x_{b},p_{b}$.

Using these fluctuation operators, we can define the covariance matrix $\Sigma^{\alpha \beta} $ as follows
\begin{equation}
\Sigma^{\alpha \beta} = \frac{\langle \{F^{\alpha},F^{\beta}\} \rangle }{2}=\frac{C^{\alpha \beta}+(C^{\alpha \beta})^T}{2},    
\end{equation}
where $C^{\alpha \beta}:= \langle F^{\alpha}F^{\beta} \rangle$ is a two-point correlation function.

The time evolution of the covariance matrix is given by $\Dot{\Sigma}=(\Dot{C}^{\alpha \beta}+(\Dot{C}^{\alpha \beta})^T)/2$, where
\begin{eqnarray}
    \Dot{C}^{\alpha \beta}=\langle \mathcal{L}[F^{\alpha}F^{\beta}] \rangle = \langle iF^\alpha [H, F^\beta] \rangle +\langle i[H,F^\alpha]  F^\beta \rangle + \langle \mathcal{D}[F^{\alpha}F^{\beta}] \rangle. \label{eq:corr}
\end{eqnarray}

Let us first solve for the term $ \langle i [H, F^\alpha] F^\beta\rangle$.
We start with the commutators of $\hat{x}_\alpha$ and $\hat{p}_\alpha$ with $H$, which are given as follows
\begin{align}
    i[H,\hat{x}_\alpha]&=\Omega \hat{p}_{\gamma \neq \alpha}/2 + 2U \hat{p}_\alpha(2\hat{n}_\alpha-1)/N \\
    i[H,\hat{p}_\alpha]&=-\Omega \hat{x}_{\gamma \neq \alpha}/2 - 2U \hat{x}_\alpha(2\hat{n}_\alpha-1)/N.
\end{align}
Making use of the above equation, the term $\langle i [H, F^\alpha] F^\beta\rangle$ can be written as $(AC)^{\alpha \beta}$ where 
\begin{equation}
    A=\begin{bmatrix}
2U x_ap_a & U (E_a+2 p_a^2) & 0 & \Omega/2 \\
-U (E_a+2x_a^2) & -2Ux_ap_a & -\Omega/2 & 0 \\
0 & \Omega/2 & 2Ux_b p_b & U (E_b+2 p_b^2) \\
-\Omega/2 & 0 & -U (E_b+2x_b^2) & -2Ux_b p_b
\end{bmatrix}.
\end{equation}
Here $E_\alpha= x_\alpha^2+p_\alpha^2$. Similarly, the term $\langle iF^\alpha [H, F^\beta] \rangle$ can be written as $(CA)^{\alpha \beta}$.

We now solve for the last term of Eq.~\eqref{eq:corr} $\mathcal{D}[F^{\alpha}F^{\beta}]$, which can be expanded as follows
\begin{equation}
    \mathcal{D}[F^{\alpha}F^{\beta}]=F^{\alpha}\mathcal{D}[F^{\beta}]+\mathcal{D}[F^{\alpha}]F^{\beta}+ \kappa (1+n_{th}) [\hat{a}\hat{b}^\dagger ,F^{\alpha}][F^{\beta},\hat{a}^\dagger \hat{b}]+ \kappa n_{th} [\hat{a}^\dagger \hat{b} ,F^{\alpha}][F^{\beta},\hat{a} \hat{b}^\dagger].\label{eq:dissipator}
\end{equation}
Let us calculate $\mathcal{D}[F^{\alpha}]$ for all $F^{\alpha}$'s:
\begin{eqnarray}
    \mathcal{D}[F^{x_a}]&=&-\kappa \hat{x}_a (1+\hat{n}_b)/N, \nonumber \\
    \mathcal{D}[F^{p_a}]&=&-\kappa \hat{p}_a (1+\hat{n}_b)/N, \nonumber \\
        \mathcal{D}[F^{x_b}]&=&\kappa \hat{x}_b \hat{n}_a/N, \nonumber \\
    \mathcal{D}[F^{p_b}]&=&\kappa \hat{p}_b \hat{n}_a/N.
\end{eqnarray}

Making use of the above equations, the term $\langle \mathcal{D}[F^{\alpha}]F^{\beta} \rangle$ in Eq.~(\ref{eq:dissipator}) can be simplified as $(QC)^{\alpha \beta}$ where 
\begin{equation}
    Q= \frac{\kappa}{4}
\begin{bmatrix}
-E_b & 0 & -2x_ax_b & -2x_a p_b \\
0 & -E_b & -2p_ax_b & -2p_a p_b \\
2x_ax_b & 2p_ax_b & E_a & 0 \\
2x_a p_b & 2p_a p_b & 0 & E_a
\end{bmatrix}.
\end{equation}

Similarly $\langle F^{\alpha} \mathcal{D}[F^{\beta}] \rangle = (CQ)^{\alpha \beta}$.

To solve for the remaining terms of Eq.~(\ref{eq:dissipator}), 
we calculate the following expressions
\begin{align}
    [\hat{a}^\dagger \hat{b} , F^{x_\alpha}] &= \frac{1}{2}(-i \hat{p}_{\gamma \neq \alpha } + \delta_{\alpha ,b } \hat{x}_a - \delta_{\alpha ,a } \hat{x}_b ), \nonumber \\
    [ \hat{a}^\dagger \hat{b} , F^{p_\alpha} ] &= \frac{1}{2} (i \hat{x}_{\gamma \neq \alpha} + \delta_{\alpha ,b} \hat{p}_a - \delta_{\alpha ,a} \hat{p}_b), \nonumber \\
    [  F^{x_\beta}, \hat{a} \hat{b}^\dagger ] &= \frac{1}{2} (i \hat{p}_{\gamma \neq \beta} + \delta_{\beta ,b} \hat{x}_a - \delta_{\beta ,a} \hat{x}_b), \nonumber \\
     [  F^{p_\beta}, \hat{a} \hat{b}^\dagger ] &= \frac{1}{2} (-i \hat{x}_{\gamma \neq \beta} + \delta_{\beta ,b} \hat{p}_a - \delta_{\beta ,a} \hat{p}_b).
\end{align}

Using the above equations, the matrix $Z=\langle \kappa (1+n_{th}) [\hat{a}\hat{b}^\dagger ,F^{\alpha}][F^{\beta},\hat{a}^\dagger \hat{b}]+ \kappa n_{th} [\hat{a}^\dagger \hat{b} ,F^{\alpha}][F^{\beta},\hat{a} \hat{b}^\dagger]\rangle$ can be written as follows
\begin{equation}
    Z=\frac{\kappa(2n_{th}+1)}{4}\begin{bmatrix}
E_b & iE_b & p_a p_b-x_ax_b  & -i(p_a p_b-x_ax_b) \\
 &  & - i(x_a p_b+x_bp_a) &  -(x_a p_b+x_bp_a)\\
 &  &  &  \\
-iE_b & E_b & i(p_a p_b-x_ax_b) & -(p_a p_b-x_ax_b) \\
 &  & -(x_a p_b+x_bp_a) & + i(x_a p_b+x_bp_a) \\
  &  &  &  \\
p_a p_b-x_ax_b & -i(p_a p_b-x_ax_b) & E_a & -iE_a \\
+ i(x_a p_b+x_bp_a) & -(x_a p_b+x_bp_a) &  &  \\
  &  &  &  \\
i(p_a p_b-x_ax_b) & -(p_a p_b-x_ax_b) & iE_a & E_a \\
-(x_a p_b+x_bp_a) & -i(x_a p_b+x_bp_a)&  & 
\end{bmatrix}.
\end{equation}

Finally, the time evolution of the covariance matrix reduces to the following form,
\begin{equation}
    \dot\Sigma= \Sigma(A^T+Q^T)+(A+Q)\Sigma + (Z+Z^T)/2.
\end{equation}

\section{Quantum and classical correlations \label{sec:entanglement}}

The correlation matrix can be used to calculate the quantum and classical correlations of the system \cite{boneberg2022quantum,mattes2023entangled}.
We time-evolve the correlation matrix and calculate the average amount of entanglement ($\varepsilon$), quantum discord ($\mathcal{D}^{a \leftarrow b}$), and classical correlations ($\mathcal{J}^{a \leftarrow b}$).

\begin{figure}[b!]
    \centering
    \includegraphics[width=1\textwidth]{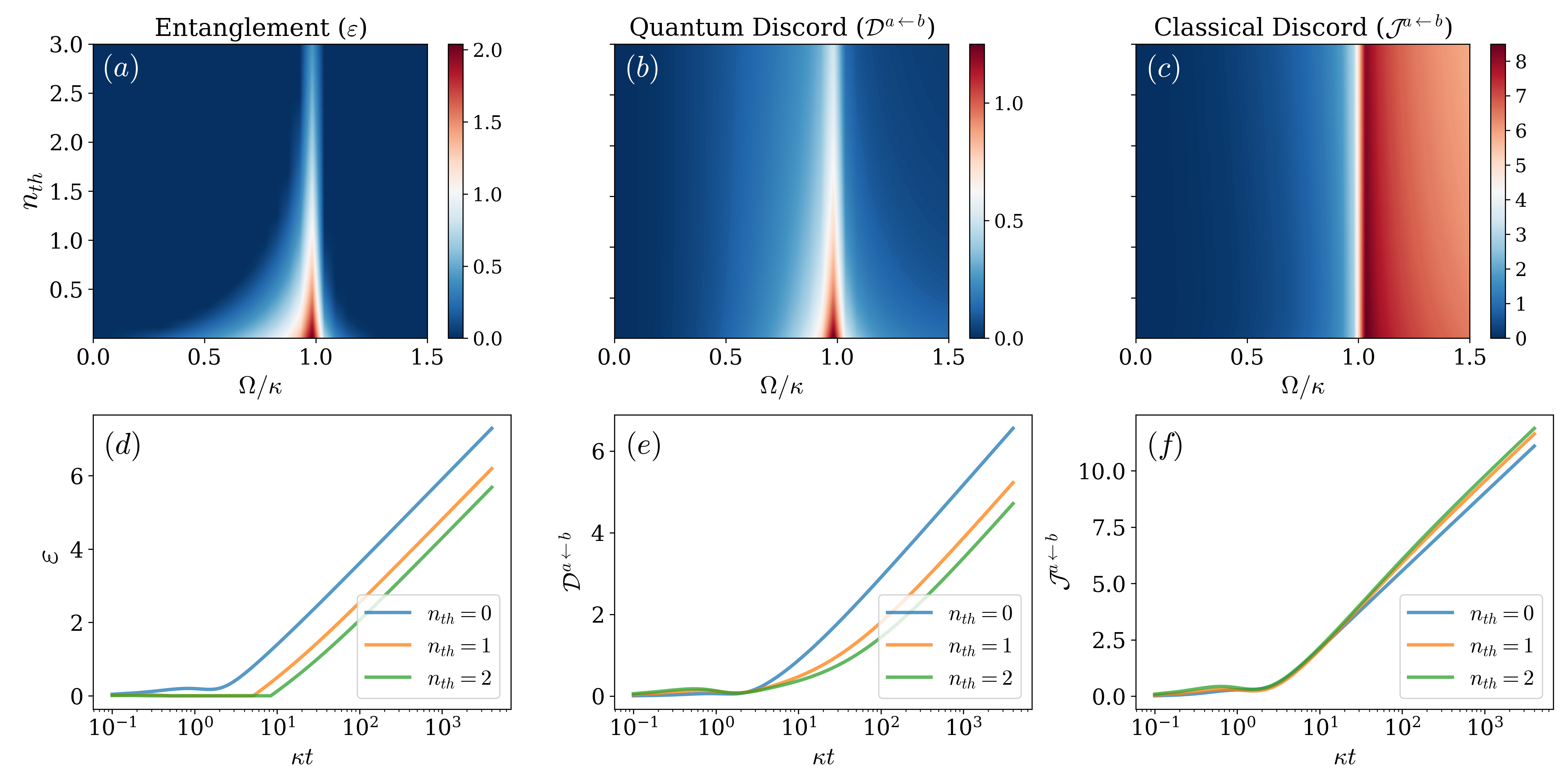}
    \caption{Dependence of time-averaged entanglement ($\varepsilon$), quantum discord ($\mathcal{D}^{a \leftarrow b}$) and classical discord ($\mathcal{J}^{a \leftarrow b}$) on $n_{th}$ for a range of $\Omega/\kappa$ values with $U=0$. Entanglement (a) and quantum discord (b) exhibit maxima at $\Omega_c/\kappa$ and decrease with increasing $n_{th}$. Panels (d) and (e) show that $\varepsilon$ and $\mathcal{D}^{a \leftarrow b}$ grow at a rate proportional to $\ln(\kappa t)$. An increase in $n_{th}$ delays the onset of quantum correlations in (d,e) but does not change their rate. (c) Classical correlations ($\mathcal{J}^{a \leftarrow b}$) also exhibit a sharp transition at $\Omega/\kappa=1$, while they increase with increasing $n_{th}$, as shown in (f). }
    \label{fig:correlations_BTC}
\end{figure}

Let us first discuss the effect of finite $n_{th}$ on the quantum and classical correlations for $U/\kappa=0$.
Both quantum and classical correlations exhibit maxima at the phase transition point $\Omega/\kappa=1$ as shown in Figs.~\ref{fig:correlations_BTC}(a,b,c).
The amount of quantum correlations (entanglement and quantum discord) decreases with increasing $n_{th}$, but the classical correlation almost remains unchanged.
To further investigate the effect of finite $n_{th}$, we investigate the spreading of correlations as the system evolves in time.
As shown in Fig.~\ref{fig:correlations_BTC}(d), the entanglement increases logarithmic in time i.e. $\varepsilon \propto \ln(\kappa t)$.
This behavior indicates the divergence of correlation at the transition point, a feature of second-order phase transition.
Interestingly, the quantum correlation between the two modes diverges even at a finite $n_{th}$.
An increase in $n_{th}$ does not affect the rate of the entanglement growth, but it induces a delay in the onset of entanglement between two modes, as shown in Fig.~\ref{fig:correlations_BTC}(d).
The higher $n_{th}$, the longer it takes for the system to get entangled, but then entanglement grows with the same logarithmic behavior.
Similar effects are observed for the quantum discord as shown in Fig.~\ref{fig:correlations_BTC}(e).
Interestingly, as shown in Fig.~\ref{fig:correlations_BTC}(f), the increase in $n_{th}$ does not affect the delay in the onset of classical correlations, but it changes the growth rate of $\mathcal{J}^{a \leftarrow b}$.
The slope of $\mathcal{J}^{a \leftarrow b}$ with respect to $\ln(\kappa t)$ increases with an increase in $n_{th}$.
This indicates that the increase in $n_{th}$ favors the higher amount of classical correlation between the two modes.

To understand the critical behavior, we look at the scaling of the order parameter $\Delta N= \vert \langle \hat{a}^\dagger \hat{a} -\hat{b}^\dagger \hat{b} \rangle\vert/2 = \vert R_a^2-R_b^2 \vert/2$ near the phase transition point $\Omega_c/\kappa=1$.
The order parameter $\Delta N\neq 0$ for $\Omega/\kappa<\Omega_c/\kappa$ and vanishes for $\Omega/\kappa>\Omega_c/\kappa$.
Following the fixed-point analysis and using the solution for $R_{a,b}$ from Eq.~(\ref{eq:mf_solutions}), the order parameter simplifies as $\Delta N = \sqrt{1-(\Omega/\kappa)^2}$ for $\Omega/\kappa<1$.
To understand the scaling behavior around the phase transition point, we use $\Omega/\kappa \rightarrow \Omega_c/\kappa-\eta$, where $\eta$ is a perturbatively small parameter ($\eta \ll 1$).
The order parameter simplifies as $\Delta N = \sqrt{1-(1-\eta)^2}\propto \eta^{1/2}$, which can be expressed as $\Delta N\propto[(\Omega_c-\Omega)/\kappa]^{1/2}$.
Thus, the system undergoes a second-order phase transition at $\Omega_c/\kappa$, characterized by the critical exponent $1/2$.

\begin{figure}[h!]
    \centering
    \includegraphics[width=1\textwidth]{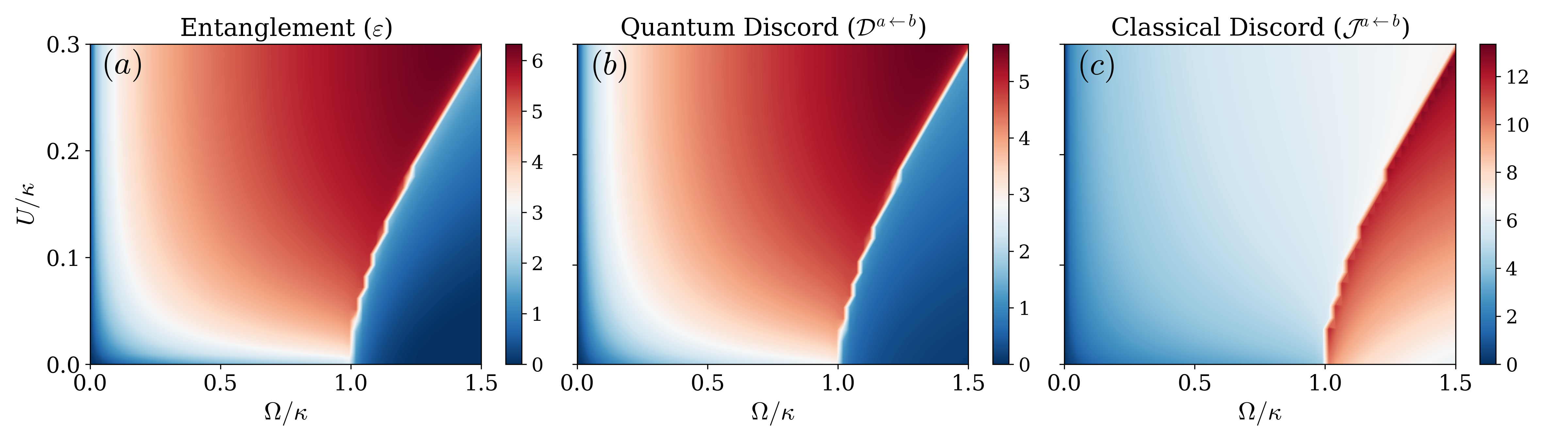}
    \caption{Panels (a), (b) and (c) depict the time-averaged entanglement ($\varepsilon$), quantum discord ($\mathcal{D}^{a \leftarrow b}$) and classical discord ($\mathcal{J}^{a \leftarrow b}$), respectively, across various values of $\Omega/\kappa$ and $U/\kappa$ with $n_{th}=0$.}
    \label{fig:correlations_U}
\end{figure}

In conclusion, both quantum and classical correlations diverge at the phase transition point $\Omega/\kappa=1$ for $U/\kappa=0$, but rapidly vanish with changes in $\Omega/\kappa$ values.

For $U/\kappa \neq 0$, the system shows a change in the entanglement between two modes at $U_c/\kappa$. 
The entanglement and quantum discord grow with increasing $\Omega/\kappa$ for $U>U_c$ and decay rapidly for $U<U_c$.
On the contrary, the classical discord shows the inverse behavior.
The classical discord shows a low amount of correlations for $U>U_c$, which increases sharply for $U<U_c$.
The entanglement saturates with time at the transition point $U_c/\kappa$.
The saturation of entanglement with time, along with the existence of a bistable regime around $U_c/\kappa$, indicates that the transition is of first order.

\section{Spin-equivalent model using the Schwinger transformation \label{sec:BTC}}

Here, we discuss a spin-equivalent representation of our model.
We use the Schwinger-Boson transformation, which is defined as follows
\begin{equation}
    \hat{S}_+=\hat{a}^\dagger \hat{b}, ~~~ \hat{S}_-=\hat{a} \hat{b}^\dagger, ~~~ \hat{S}_z=(\hat{a}^\dagger \hat{a} - \hat{b}^\dagger \hat{b})/2,
\end{equation}
where the total spin of the system is described by $\hat{S}=(\hat{a}^\dagger \hat{a} + \hat{b}^\dagger \hat{b})/2$, with $\hat{S}_{x,y,z}$ being total angular momentum operators and $\hat{S}_\pm=\hat{S}_x\pm i\hat{S}_y$.
Using the above transformation, Eq.~(\ref{eq:ME_ab}) can be re-written as follows,
\begin{align}
    \dot \rho = &-i [\Omega \hat{S}_x + \frac{2U}{S} (\hat{S}^2+\hat{S}_z^2-\hat{S}), \rho] +(1+n_{th})\frac{\kappa}{S}\mathcal{D}[\hat{S}_-] \rho + n_{th}\frac{\kappa}{S}\mathcal{D}[\hat{S}_+] \rho,
    \label{eq:BTC_themal}
\end{align}
where $S=\langle \hat{S} \rangle=N/2$.
The above equation reduces to the well-known boundary time crystal (BTC) in the limit $U/\kappa=0$ and $n_{th}=0$.
The BTC also exhibits a phase transition between a melted phase ($\Omega/\kappa<1$) and a time crystal phase ($\Omega/\kappa>1$) at $\Omega/\kappa=1$.

\begin{figure}[t!]
    \centering
    \includegraphics[width=0.8\linewidth]{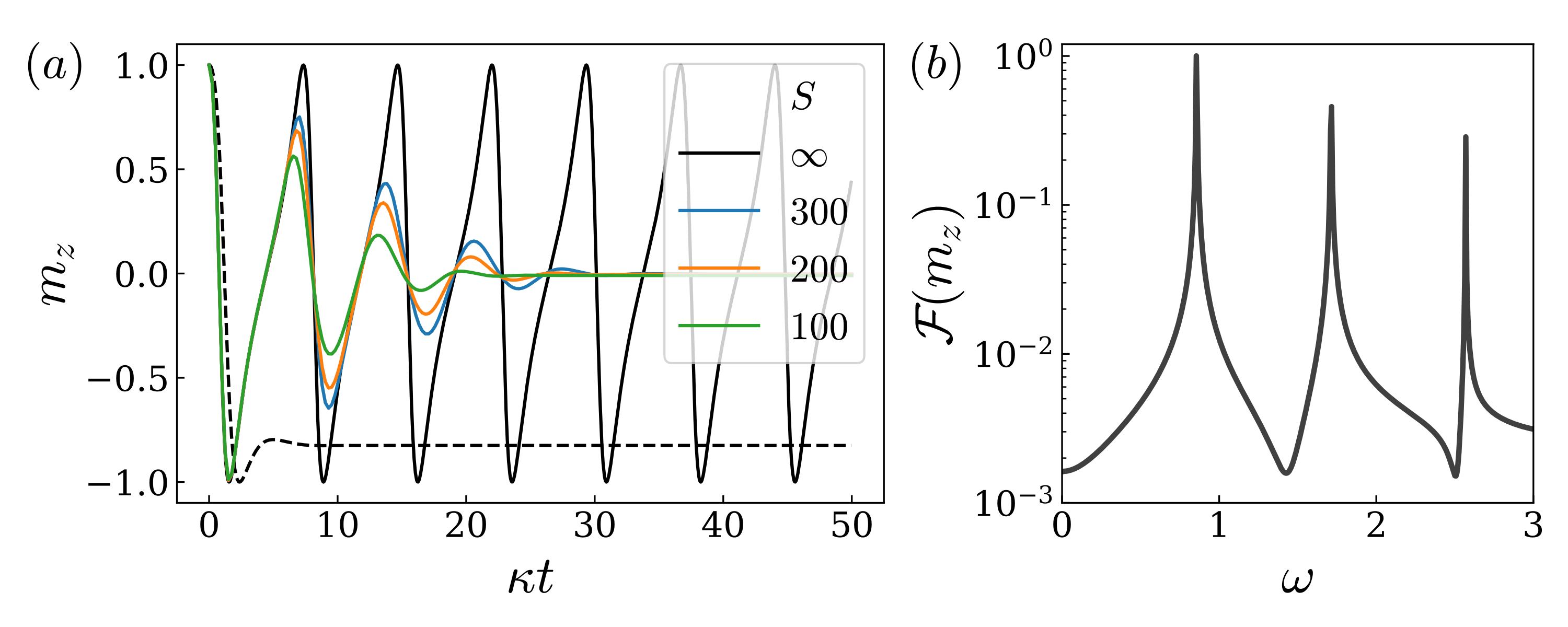}
    \caption{(a) Time evolution of the observable $m_z=\langle S_z\rangle/S$ for $\Omega/\kappa=1.45$ (solid lines) and $\Omega/\kappa=0.8$ (dashed line) for a fixed value of $U/\kappa=0.25$. The spin system attains a stationary steady state for $U>U_c$,  indicated by the dashed line, and exhibits persistent oscillations for $U<U_c$ (depicted by the solid line) in the thermodynamic limit $S\rightarrow \infty$. (b) Fourier transform of the time evolution of $m_z$ for $S\rightarrow \infty$. }
    \label{fig:BTC}
\end{figure}

Apart from the similarities, for $U/\kappa\neq 0$, the above spin model features distinct phases than the corresponding open Bose-Hubbard model given by Eq.~(\ref{eq:ME_ab}).
The spin model exhibits a melted phase for $U>U_c$ and a time crystal phase for $U<U_c$.
This is in contrast to the behavior of the Bose-Hubbard model discussed in the previous sections, where the bosonic system exists in a time crystal phase TC2 for $U>U_c$ and a quasi-periodic dynamics for $U<U_c$.
This is due to the fact that the sub-systems $a$ and $b$ oscillate with constant distinct radii for $U>U_c$ while keeping the $S_z=(a^\dagger a - b^\dagger b)/2$ constant in time.
For $U<U_c$, the sub-systems $a$ and $b$ exhibit quasi-periodic dynamics such that the corresponding spin operator system attains a periodic limit cycle phase.
This behavior is shown in Fig.~\ref{fig:BTC} (a), where the observable $m_z=\langle S_z\rangle/S$ oscillates for the  $U<U_c$ and settles down to a time-invariant fixed point for $U>U_c$ in the thermodynamic limit of system size $S\rightarrow \infty$.

\begin{thebibliography}{69}%
\makeatletter
\providecommand \@ifxundefined [1]{%
 \@ifx{#1\undefined}
}%
\providecommand \@ifnum [1]{%
 \ifnum #1\expandafter \@firstoftwo
 \else \expandafter \@secondoftwo
 \fi
}%
\providecommand \@ifx [1]{%
 \ifx #1\expandafter \@firstoftwo
 \else \expandafter \@secondoftwo
 \fi
}%
\providecommand \natexlab [1]{#1}%
\providecommand \enquote  [1]{``#1''}%
\providecommand \bibnamefont  [1]{#1}%
\providecommand \bibfnamefont [1]{#1}%
\providecommand \citenamefont [1]{#1}%
\providecommand \href@noop [0]{\@secondoftwo}%
\providecommand \href [0]{\begingroup \@sanitize@url \@href}%
\providecommand \@href[1]{\@@startlink{#1}\@@href}%
\providecommand \@@href[1]{\endgroup#1\@@endlink}%
\providecommand \@sanitize@url [0]{\catcode `\\12\catcode `\$12\catcode `\&12\catcode `\#12\catcode `\^12\catcode `\_12\catcode `\%12\relax}%
\providecommand \@@startlink[1]{}%
\providecommand \@@endlink[0]{}%
\providecommand \url  [0]{\begingroup\@sanitize@url \@url }%
\providecommand \@url [1]{\endgroup\@href {#1}{\urlprefix }}%
\providecommand \urlprefix  [0]{URL }%
\providecommand \Eprint [0]{\href }%
\providecommand \doibase [0]{https://doi.org/}%
\providecommand \selectlanguage [0]{\@gobble}%
\providecommand \bibinfo  [0]{\@secondoftwo}%
\providecommand \bibfield  [0]{\@secondoftwo}%
\providecommand \translation [1]{[#1]}%
\providecommand \BibitemOpen [0]{}%
\providecommand \bibitemStop [0]{}%
\providecommand \bibitemNoStop [0]{.\EOS\space}%
\providecommand \EOS [0]{\spacefactor3000\relax}%
\providecommand \BibitemShut  [1]{\csname bibitem#1\endcsname}%
\let\auto@bib@innerbib\@empty
\bibitem [{\citenamefont {S{\o}rensen}\ \emph {et~al.}(2001)\citenamefont {S{\o}rensen}, \citenamefont {Duan}, \citenamefont {Cirac},\ and\ \citenamefont {Zoller}}]{sorensen2001many}%
  \BibitemOpen
  \bibfield  {author} {\bibinfo {author} {\bibfnamefont {A.}~\bibnamefont {S{\o}rensen}}, \bibinfo {author} {\bibfnamefont {L.-M.}\ \bibnamefont {Duan}}, \bibinfo {author} {\bibfnamefont {J.~I.}\ \bibnamefont {Cirac}},\ and\ \bibinfo {author} {\bibfnamefont {P.}~\bibnamefont {Zoller}},\ }\bibfield  {title} {\bibinfo {title} {Many-particle entanglement with {B}ose--{E}instein condensates},\ }\href {https://doi.org/10.1038/35051038} {\bibfield  {journal} {\bibinfo  {journal} {Nature}\ }\textbf {\bibinfo {volume} {409}},\ \bibinfo {pages} {63} (\bibinfo {year} {2001})}\BibitemShut {NoStop}%
\bibitem [{\citenamefont {Fazio}\ \emph {et~al.}(2024)\citenamefont {Fazio}, \citenamefont {Keeling}, \citenamefont {Mazza},\ and\ \citenamefont {Schirò}}]{fazio2024manybodyopenquantumsystems}%
  \BibitemOpen
  \bibfield  {author} {\bibinfo {author} {\bibfnamefont {R.}~\bibnamefont {Fazio}}, \bibinfo {author} {\bibfnamefont {J.}~\bibnamefont {Keeling}}, \bibinfo {author} {\bibfnamefont {L.}~\bibnamefont {Mazza}},\ and\ \bibinfo {author} {\bibfnamefont {M.}~\bibnamefont {Schirò}},\ }\href {https://arxiv.org/abs/2409.10300} {\bibinfo {title} {Many-body open quantum systems}} (\bibinfo {year} {2024}),\ \Eprint {https://arxiv.org/abs/2409.10300} {arXiv:2409.10300 [quant-ph]} \BibitemShut {NoStop}%
\bibitem [{\citenamefont {Korbicz}\ \emph {et~al.}(2005)\citenamefont {Korbicz}, \citenamefont {Cirac},\ and\ \citenamefont {Lewenstein}}]{korbicz2005spin}%
  \BibitemOpen
  \bibfield  {author} {\bibinfo {author} {\bibfnamefont {J.~K.}\ \bibnamefont {Korbicz}}, \bibinfo {author} {\bibfnamefont {J.~I.}\ \bibnamefont {Cirac}},\ and\ \bibinfo {author} {\bibfnamefont {M.}~\bibnamefont {Lewenstein}},\ }\bibfield  {title} {\bibinfo {title} {Spin squeezing inequalities and entanglement of $n$ qubit states},\ }\href {https://doi.org/10.1103/PhysRevLett.95.120502} {\bibfield  {journal} {\bibinfo  {journal} {Phys. Rev. Lett.}\ }\textbf {\bibinfo {volume} {95}},\ \bibinfo {pages} {120502} (\bibinfo {year} {2005})}\BibitemShut {NoStop}%
\bibitem [{\citenamefont {Griffin}\ \emph {et~al.}(1996)\citenamefont {Griffin}, \citenamefont {Snoke},\ and\ \citenamefont {Stringari}}]{griffin1996bose}%
  \BibitemOpen
  \bibfield  {author} {\bibinfo {author} {\bibfnamefont {A.}~\bibnamefont {Griffin}}, \bibinfo {author} {\bibfnamefont {D.~W.}\ \bibnamefont {Snoke}},\ and\ \bibinfo {author} {\bibfnamefont {S.}~\bibnamefont {Stringari}},\ }\href@noop {} {\emph {\bibinfo {title} {{B}ose-{E}instein condensation}}}\ (\bibinfo  {publisher} {Cambridge University Press},\ \bibinfo {year} {1996})\BibitemShut {NoStop}%
\bibitem [{\citenamefont {Hofer}\ \emph {et~al.}(2012)\citenamefont {Hofer}, \citenamefont {Bruder},\ and\ \citenamefont {Stojanovi\ifmmode~\acute{c}\else \'{c}\fi{}}}]{hofer2012superfluid}%
  \BibitemOpen
  \bibfield  {author} {\bibinfo {author} {\bibfnamefont {P.~P.}\ \bibnamefont {Hofer}}, \bibinfo {author} {\bibfnamefont {C.}~\bibnamefont {Bruder}},\ and\ \bibinfo {author} {\bibfnamefont {V.~M.}\ \bibnamefont {Stojanovi\ifmmode~\acute{c}\else \'{c}\fi{}}},\ }\bibfield  {title} {\bibinfo {title} {Superfluid drag of two-species {B}ose-{E}instein condensates in optical lattices},\ }\href {https://doi.org/10.1103/PhysRevA.86.033627} {\bibfield  {journal} {\bibinfo  {journal} {Phys. Rev. A}\ }\textbf {\bibinfo {volume} {86}},\ \bibinfo {pages} {033627} (\bibinfo {year} {2012})}\BibitemShut {NoStop}%
\bibitem [{\citenamefont {Lode}\ and\ \citenamefont {Bruder}(2017)}]{lode2017fragmented}%
  \BibitemOpen
  \bibfield  {author} {\bibinfo {author} {\bibfnamefont {A.~U.~J.}\ \bibnamefont {Lode}}\ and\ \bibinfo {author} {\bibfnamefont {C.}~\bibnamefont {Bruder}},\ }\bibfield  {title} {\bibinfo {title} {Fragmented superradiance of a {B}ose-{E}instein condensate in an optical cavity},\ }\href {https://doi.org/10.1103/PhysRevLett.118.013603} {\bibfield  {journal} {\bibinfo  {journal} {Phys. Rev. Lett.}\ }\textbf {\bibinfo {volume} {118}},\ \bibinfo {pages} {013603} (\bibinfo {year} {2017})}\BibitemShut {NoStop}%
\bibitem [{\citenamefont {Autti}\ \emph {et~al.}(2018)\citenamefont {Autti}, \citenamefont {Eltsov},\ and\ \citenamefont {Volovik}}]{PhysRevLett.120.215301}%
  \BibitemOpen
  \bibfield  {author} {\bibinfo {author} {\bibfnamefont {S.}~\bibnamefont {Autti}}, \bibinfo {author} {\bibfnamefont {V.~B.}\ \bibnamefont {Eltsov}},\ and\ \bibinfo {author} {\bibfnamefont {G.~E.}\ \bibnamefont {Volovik}},\ }\bibfield  {title} {\bibinfo {title} {Observation of a time quasicrystal and its transition to a superfluid time crystal},\ }\href {https://doi.org/10.1103/PhysRevLett.120.215301} {\bibfield  {journal} {\bibinfo  {journal} {Phys. Rev. Lett.}\ }\textbf {\bibinfo {volume} {120}},\ \bibinfo {pages} {215301} (\bibinfo {year} {2018})}\BibitemShut {NoStop}%
\bibitem [{\citenamefont {Autti}\ \emph {et~al.}(2021)\citenamefont {Autti}, \citenamefont {Heikkinen}, \citenamefont {M{\"a}kinen}, \citenamefont {Volovik}, \citenamefont {Zavjalov},\ and\ \citenamefont {Eltsov}}]{autti2021ac}%
  \BibitemOpen
  \bibfield  {author} {\bibinfo {author} {\bibfnamefont {S.}~\bibnamefont {Autti}}, \bibinfo {author} {\bibfnamefont {P.~J.}\ \bibnamefont {Heikkinen}}, \bibinfo {author} {\bibfnamefont {J.~T.}\ \bibnamefont {M{\"a}kinen}}, \bibinfo {author} {\bibfnamefont {G.~E.}\ \bibnamefont {Volovik}}, \bibinfo {author} {\bibfnamefont {V.~V.}\ \bibnamefont {Zavjalov}},\ and\ \bibinfo {author} {\bibfnamefont {V.~B.}\ \bibnamefont {Eltsov}},\ }\bibfield  {title} {\bibinfo {title} {{AC Josephson} effect between two superfluid time crystals},\ }\href {https://doi.org/10.1038/s41563-020-0780-y} {\bibfield  {journal} {\bibinfo  {journal} {Nat. Mater.}\ }\textbf {\bibinfo {volume} {20}},\ \bibinfo {pages} {171} (\bibinfo {year} {2021})}\BibitemShut {NoStop}%
\bibitem [{\citenamefont {Ke\ss{}ler}\ \emph {et~al.}(2021)\citenamefont {Ke\ss{}ler}, \citenamefont {Kongkhambut}, \citenamefont {Georges}, \citenamefont {Mathey}, \citenamefont {Cosme},\ and\ \citenamefont {Hemmerich}}]{kessler2021observation}%
  \BibitemOpen
  \bibfield  {author} {\bibinfo {author} {\bibfnamefont {H.}~\bibnamefont {Ke\ss{}ler}}, \bibinfo {author} {\bibfnamefont {P.}~\bibnamefont {Kongkhambut}}, \bibinfo {author} {\bibfnamefont {C.}~\bibnamefont {Georges}}, \bibinfo {author} {\bibfnamefont {L.}~\bibnamefont {Mathey}}, \bibinfo {author} {\bibfnamefont {J.~G.}\ \bibnamefont {Cosme}},\ and\ \bibinfo {author} {\bibfnamefont {A.}~\bibnamefont {Hemmerich}},\ }\bibfield  {title} {\bibinfo {title} {Observation of a dissipative time crystal},\ }\href {https://doi.org/10.1103/PhysRevLett.127.043602} {\bibfield  {journal} {\bibinfo  {journal} {Phys. Rev. Lett.}\ }\textbf {\bibinfo {volume} {127}},\ \bibinfo {pages} {043602} (\bibinfo {year} {2021})}\BibitemShut {NoStop}%
\bibitem [{\citenamefont {Smerzi}\ \emph {et~al.}(1997)\citenamefont {Smerzi}, \citenamefont {Fantoni}, \citenamefont {Giovanazzi},\ and\ \citenamefont {Shenoy}}]{PhysRevLett.79.4950}%
  \BibitemOpen
  \bibfield  {author} {\bibinfo {author} {\bibfnamefont {A.}~\bibnamefont {Smerzi}}, \bibinfo {author} {\bibfnamefont {S.}~\bibnamefont {Fantoni}}, \bibinfo {author} {\bibfnamefont {S.}~\bibnamefont {Giovanazzi}},\ and\ \bibinfo {author} {\bibfnamefont {S.~R.}\ \bibnamefont {Shenoy}},\ }\bibfield  {title} {\bibinfo {title} {Quantum coherent atomic tunneling between two trapped {B}ose-{E}instein condensates},\ }\href {https://doi.org/10.1103/PhysRevLett.79.4950} {\bibfield  {journal} {\bibinfo  {journal} {Phys. Rev. Lett.}\ }\textbf {\bibinfo {volume} {79}},\ \bibinfo {pages} {4950} (\bibinfo {year} {1997})}\BibitemShut {NoStop}%
\bibitem [{\citenamefont {Orzel}\ \emph {et~al.}(2001)\citenamefont {Orzel}, \citenamefont {Tuchman}, \citenamefont {Fenselau}, \citenamefont {Yasuda},\ and\ \citenamefont {Kasevich}}]{orzel2001squeezed}%
  \BibitemOpen
  \bibfield  {author} {\bibinfo {author} {\bibfnamefont {C.}~\bibnamefont {Orzel}}, \bibinfo {author} {\bibfnamefont {A.}~\bibnamefont {Tuchman}}, \bibinfo {author} {\bibfnamefont {M.}~\bibnamefont {Fenselau}}, \bibinfo {author} {\bibfnamefont {M.}~\bibnamefont {Yasuda}},\ and\ \bibinfo {author} {\bibfnamefont {M.}~\bibnamefont {Kasevich}},\ }\bibfield  {title} {\bibinfo {title} {Squeezed states in a {B}ose-{E}instein condensate},\ }\href {https://doi.org/10.1126/science.1058149} {\bibfield  {journal} {\bibinfo  {journal} {Science}\ }\textbf {\bibinfo {volume} {291}},\ \bibinfo {pages} {2386} (\bibinfo {year} {2001})}\BibitemShut {NoStop}%
\bibitem [{\citenamefont {Est\`eve}\ \emph {et~al.}(2008)\citenamefont {Est\`eve}, \citenamefont {Gross}, \citenamefont {Weller}, \citenamefont {Giovanazzi},\ and\ \citenamefont {Oberthaler}}]{esteve2008squeezing}%
  \BibitemOpen
  \bibfield  {author} {\bibinfo {author} {\bibfnamefont {J.}~\bibnamefont {Est\`eve}}, \bibinfo {author} {\bibfnamefont {C.}~\bibnamefont {Gross}}, \bibinfo {author} {\bibfnamefont {A.}~\bibnamefont {Weller}}, \bibinfo {author} {\bibfnamefont {S.}~\bibnamefont {Giovanazzi}},\ and\ \bibinfo {author} {\bibfnamefont {M.~K.}\ \bibnamefont {Oberthaler}},\ }\bibfield  {title} {\bibinfo {title} {Squeezing and entanglement in a {B}ose--{E}instein condensate},\ }\href {https://doi.org/10.1038/nature07332} {\bibfield  {journal} {\bibinfo  {journal} {Nature}\ }\textbf {\bibinfo {volume} {455}},\ \bibinfo {pages} {1216} (\bibinfo {year} {2008})}\BibitemShut {NoStop}%
\bibitem [{\citenamefont {Iemini}\ \emph {et~al.}(2018)\citenamefont {Iemini}, \citenamefont {Russomanno}, \citenamefont {Keeling}, \citenamefont {Schir\`o}, \citenamefont {Dalmonte},\ and\ \citenamefont {Fazio}}]{iemini2018boundary}%
  \BibitemOpen
  \bibfield  {author} {\bibinfo {author} {\bibfnamefont {F.}~\bibnamefont {Iemini}}, \bibinfo {author} {\bibfnamefont {A.}~\bibnamefont {Russomanno}}, \bibinfo {author} {\bibfnamefont {J.}~\bibnamefont {Keeling}}, \bibinfo {author} {\bibfnamefont {M.}~\bibnamefont {Schir\`o}}, \bibinfo {author} {\bibfnamefont {M.}~\bibnamefont {Dalmonte}},\ and\ \bibinfo {author} {\bibfnamefont {R.}~\bibnamefont {Fazio}},\ }\bibfield  {title} {\bibinfo {title} {Boundary time crystals},\ }\href {https://doi.org/10.1103/PhysRevLett.121.035301} {\bibfield  {journal} {\bibinfo  {journal} {Phys. Rev. Lett.}\ }\textbf {\bibinfo {volume} {121}},\ \bibinfo {pages} {035301} (\bibinfo {year} {2018})}\BibitemShut {NoStop}%
\bibitem [{\citenamefont {Tucker}\ \emph {et~al.}(2018)\citenamefont {Tucker}, \citenamefont {Zhu}, \citenamefont {Lewis-Swan}, \citenamefont {Marino}, \citenamefont {Jimenez}, \citenamefont {Restrepo},\ and\ \citenamefont {Rey}}]{tucker2018shattered}%
  \BibitemOpen
  \bibfield  {author} {\bibinfo {author} {\bibfnamefont {K.}~\bibnamefont {Tucker}}, \bibinfo {author} {\bibfnamefont {B.}~\bibnamefont {Zhu}}, \bibinfo {author} {\bibfnamefont {R.~J.}\ \bibnamefont {Lewis-Swan}}, \bibinfo {author} {\bibfnamefont {J.}~\bibnamefont {Marino}}, \bibinfo {author} {\bibfnamefont {F.}~\bibnamefont {Jimenez}}, \bibinfo {author} {\bibfnamefont {J.~G.}\ \bibnamefont {Restrepo}},\ and\ \bibinfo {author} {\bibfnamefont {A.~M.}\ \bibnamefont {Rey}},\ }\bibfield  {title} {\bibinfo {title} {Shattered time: can a dissipative time crystal survive many-body correlations?},\ }\href {https://doi.org/10.1088/1367-2630/aaf18b} {\bibfield  {journal} {\bibinfo  {journal} {New J. Phys.}\ }\textbf {\bibinfo {volume} {20}},\ \bibinfo {pages} {123003} (\bibinfo {year} {2018})}\BibitemShut {NoStop}%
\bibitem [{\citenamefont {Bu{\v{c}}a}\ \emph {et~al.}(2019)\citenamefont {Bu{\v{c}}a}, \citenamefont {Tindall},\ and\ \citenamefont {Jaksch}}]{buca2019non}%
  \BibitemOpen
  \bibfield  {author} {\bibinfo {author} {\bibfnamefont {B.}~\bibnamefont {Bu{\v{c}}a}}, \bibinfo {author} {\bibfnamefont {J.}~\bibnamefont {Tindall}},\ and\ \bibinfo {author} {\bibfnamefont {D.}~\bibnamefont {Jaksch}},\ }\bibfield  {title} {\bibinfo {title} {Non-stationary coherent quantum many-body dynamics through dissipation},\ }\href {https://doi.org/10.1038/s41467-019-09757-y} {\bibfield  {journal} {\bibinfo  {journal} {Nat. Commun.}\ }\textbf {\bibinfo {volume} {10}},\ \bibinfo {pages} {1730} (\bibinfo {year} {2019})}\BibitemShut {NoStop}%
\bibitem [{\citenamefont {Booker}\ \emph {et~al.}(2020)\citenamefont {Booker}, \citenamefont {Buča},\ and\ \citenamefont {Jaksch}}]{Booker_2020}%
  \BibitemOpen
  \bibfield  {author} {\bibinfo {author} {\bibfnamefont {C.}~\bibnamefont {Booker}}, \bibinfo {author} {\bibfnamefont {B.}~\bibnamefont {Buča}},\ and\ \bibinfo {author} {\bibfnamefont {D.}~\bibnamefont {Jaksch}},\ }\bibfield  {title} {\bibinfo {title} {Non-stationarity and dissipative time crystals: spectral properties and finite-size effects},\ }\href {https://doi.org/10.1088/1367-2630/ababc4} {\bibfield  {journal} {\bibinfo  {journal} {New J. Phys.}\ }\textbf {\bibinfo {volume} {22}},\ \bibinfo {pages} {085007} (\bibinfo {year} {2020})}\BibitemShut {NoStop}%
\bibitem [{\citenamefont {Bu\ifmmode~\check{c}\else \v{c}\fi{}a}\ and\ \citenamefont {Jaksch}(2019)}]{PhysRevLett.123.260401}%
  \BibitemOpen
  \bibfield  {author} {\bibinfo {author} {\bibfnamefont {B.}~\bibnamefont {Bu\ifmmode~\check{c}\else \v{c}\fi{}a}}\ and\ \bibinfo {author} {\bibfnamefont {D.}~\bibnamefont {Jaksch}},\ }\bibfield  {title} {\bibinfo {title} {Dissipation induced nonstationarity in a quantum gas},\ }\href {https://doi.org/10.1103/PhysRevLett.123.260401} {\bibfield  {journal} {\bibinfo  {journal} {Phys. Rev. Lett.}\ }\textbf {\bibinfo {volume} {123}},\ \bibinfo {pages} {260401} (\bibinfo {year} {2019})}\BibitemShut {NoStop}%
\bibitem [{\citenamefont {Zhu}\ \emph {et~al.}(2019)\citenamefont {Zhu}, \citenamefont {Marino}, \citenamefont {Yao}, \citenamefont {Lukin},\ and\ \citenamefont {Demler}}]{zhu2019dicke}%
  \BibitemOpen
  \bibfield  {author} {\bibinfo {author} {\bibfnamefont {B.}~\bibnamefont {Zhu}}, \bibinfo {author} {\bibfnamefont {J.}~\bibnamefont {Marino}}, \bibinfo {author} {\bibfnamefont {N.~Y.}\ \bibnamefont {Yao}}, \bibinfo {author} {\bibfnamefont {M.~D.}\ \bibnamefont {Lukin}},\ and\ \bibinfo {author} {\bibfnamefont {E.~A.}\ \bibnamefont {Demler}},\ }\bibfield  {title} {\bibinfo {title} {Dicke time crystals in driven-dissipative quantum many-body systems},\ }\href {https://doi.org/10.1088/1367-2630/ab2afe} {\bibfield  {journal} {\bibinfo  {journal} {New J. Phys.}\ }\textbf {\bibinfo {volume} {21}},\ \bibinfo {pages} {073028} (\bibinfo {year} {2019})}\BibitemShut {NoStop}%
\bibitem [{\citenamefont {Lled\'o}\ \emph {et~al.}(2019)\citenamefont {Lled\'o}, \citenamefont {Mavrogordatos},\ and\ \citenamefont {Szyma\ifmmode~\acute{n}\else \'{n}\fi{}ska}}]{lledo2019driven}%
  \BibitemOpen
  \bibfield  {author} {\bibinfo {author} {\bibfnamefont {C.}~\bibnamefont {Lled\'o}}, \bibinfo {author} {\bibfnamefont {T.~K.}\ \bibnamefont {Mavrogordatos}},\ and\ \bibinfo {author} {\bibfnamefont {M.~H.}\ \bibnamefont {Szyma\ifmmode~\acute{n}\else \'{n}\fi{}ska}},\ }\bibfield  {title} {\bibinfo {title} {Driven {B}ose-{H}ubbard dimer under nonlocal dissipation: A bistable time crystal},\ }\href {https://doi.org/10.1103/PhysRevB.100.054303} {\bibfield  {journal} {\bibinfo  {journal} {Phys. Rev. B}\ }\textbf {\bibinfo {volume} {100}},\ \bibinfo {pages} {054303} (\bibinfo {year} {2019})}\BibitemShut {NoStop}%
\bibitem [{\citenamefont {Pizzi}\ \emph {et~al.}(2019)\citenamefont {Pizzi}, \citenamefont {Knolle},\ and\ \citenamefont {Nunnenkamp}}]{pizzi2019periodn}%
  \BibitemOpen
  \bibfield  {author} {\bibinfo {author} {\bibfnamefont {A.}~\bibnamefont {Pizzi}}, \bibinfo {author} {\bibfnamefont {J.}~\bibnamefont {Knolle}},\ and\ \bibinfo {author} {\bibfnamefont {A.}~\bibnamefont {Nunnenkamp}},\ }\bibfield  {title} {\bibinfo {title} {Period-$n$ discrete time crystals and quasicrystals with ultracold bosons},\ }\href {https://doi.org/10.1103/PhysRevLett.123.150601} {\bibfield  {journal} {\bibinfo  {journal} {Phys. Rev. Lett.}\ }\textbf {\bibinfo {volume} {123}},\ \bibinfo {pages} {150601} (\bibinfo {year} {2019})}\BibitemShut {NoStop}%
\bibitem [{\citenamefont {Pizzi}\ \emph {et~al.}(2021)\citenamefont {Pizzi}, \citenamefont {Nunnenkamp},\ and\ \citenamefont {Knolle}}]{pizzi2021bistability}%
  \BibitemOpen
  \bibfield  {author} {\bibinfo {author} {\bibfnamefont {A.}~\bibnamefont {Pizzi}}, \bibinfo {author} {\bibfnamefont {A.}~\bibnamefont {Nunnenkamp}},\ and\ \bibinfo {author} {\bibfnamefont {J.}~\bibnamefont {Knolle}},\ }\bibfield  {title} {\bibinfo {title} {Bistability and time crystals in long-ranged directed percolation},\ }\href {https://doi.org/10.1038/s41467-021-21259-4} {\bibfield  {journal} {\bibinfo  {journal} {Nat. Commun.}\ }\textbf {\bibinfo {volume} {12}},\ \bibinfo {pages} {1061} (\bibinfo {year} {2021})}\BibitemShut {NoStop}%
\bibitem [{\citenamefont {Seibold}\ \emph {et~al.}(2020)\citenamefont {Seibold}, \citenamefont {Rota},\ and\ \citenamefont {Savona}}]{seibold2020dissipative}%
  \BibitemOpen
  \bibfield  {author} {\bibinfo {author} {\bibfnamefont {K.}~\bibnamefont {Seibold}}, \bibinfo {author} {\bibfnamefont {R.}~\bibnamefont {Rota}},\ and\ \bibinfo {author} {\bibfnamefont {V.}~\bibnamefont {Savona}},\ }\bibfield  {title} {\bibinfo {title} {Dissipative time crystal in an asymmetric nonlinear photonic dimer},\ }\href {https://doi.org/10.1103/PhysRevA.101.033839} {\bibfield  {journal} {\bibinfo  {journal} {Phys. Rev. A}\ }\textbf {\bibinfo {volume} {101}},\ \bibinfo {pages} {033839} (\bibinfo {year} {2020})}\BibitemShut {NoStop}%
\bibitem [{\citenamefont {Prazeres}\ \emph {et~al.}(2021)\citenamefont {Prazeres}, \citenamefont {Souza},\ and\ \citenamefont {Iemini}}]{prazeres2021boundary}%
  \BibitemOpen
  \bibfield  {author} {\bibinfo {author} {\bibfnamefont {L.~F.~d.}\ \bibnamefont {Prazeres}}, \bibinfo {author} {\bibfnamefont {L.~d.~S.}\ \bibnamefont {Souza}},\ and\ \bibinfo {author} {\bibfnamefont {F.}~\bibnamefont {Iemini}},\ }\bibfield  {title} {\bibinfo {title} {Boundary time crystals in collective $d$-level systems},\ }\href {https://doi.org/10.1103/PhysRevB.103.184308} {\bibfield  {journal} {\bibinfo  {journal} {Phys. Rev. B}\ }\textbf {\bibinfo {volume} {103}},\ \bibinfo {pages} {184308} (\bibinfo {year} {2021})}\BibitemShut {NoStop}%
\bibitem [{\citenamefont {Piccitto}\ \emph {et~al.}(2021)\citenamefont {Piccitto}, \citenamefont {Wauters}, \citenamefont {Nori},\ and\ \citenamefont {Shammah}}]{piccitto2021symmetries}%
  \BibitemOpen
  \bibfield  {author} {\bibinfo {author} {\bibfnamefont {G.}~\bibnamefont {Piccitto}}, \bibinfo {author} {\bibfnamefont {M.}~\bibnamefont {Wauters}}, \bibinfo {author} {\bibfnamefont {F.}~\bibnamefont {Nori}},\ and\ \bibinfo {author} {\bibfnamefont {N.}~\bibnamefont {Shammah}},\ }\bibfield  {title} {\bibinfo {title} {Symmetries and conserved quantities of boundary time crystals in generalized spin models},\ }\href {https://doi.org/10.1103/PhysRevB.104.014307} {\bibfield  {journal} {\bibinfo  {journal} {Phys. Rev. B}\ }\textbf {\bibinfo {volume} {104}},\ \bibinfo {pages} {014307} (\bibinfo {year} {2021})}\BibitemShut {NoStop}%
\bibitem [{\citenamefont {Carollo}\ and\ \citenamefont {Lesanovsky}(2022)}]{carollo2022exact}%
  \BibitemOpen
  \bibfield  {author} {\bibinfo {author} {\bibfnamefont {F.}~\bibnamefont {Carollo}}\ and\ \bibinfo {author} {\bibfnamefont {I.}~\bibnamefont {Lesanovsky}},\ }\bibfield  {title} {\bibinfo {title} {Exact solution of a boundary time-crystal phase transition: Time-translation symmetry breaking and non-{M}arkovian dynamics of correlations},\ }\href {https://doi.org/10.1103/PhysRevA.105.L040202} {\bibfield  {journal} {\bibinfo  {journal} {Phys. Rev. A}\ }\textbf {\bibinfo {volume} {105}},\ \bibinfo {pages} {L040202} (\bibinfo {year} {2022})}\BibitemShut {NoStop}%
\bibitem [{\citenamefont {Krishna}\ \emph {et~al.}(2023{\natexlab{a}})\citenamefont {Krishna}, \citenamefont {Solanki}, \citenamefont {Hajdu\ifmmode~\check{s}\else \v{s}\fi{}ek},\ and\ \citenamefont {Vinjanampathy}}]{krishna2022measurement}%
  \BibitemOpen
  \bibfield  {author} {\bibinfo {author} {\bibfnamefont {M.}~\bibnamefont {Krishna}}, \bibinfo {author} {\bibfnamefont {P.}~\bibnamefont {Solanki}}, \bibinfo {author} {\bibfnamefont {M.}~\bibnamefont {Hajdu\ifmmode~\check{s}\else \v{s}\fi{}ek}},\ and\ \bibinfo {author} {\bibfnamefont {S.}~\bibnamefont {Vinjanampathy}},\ }\bibfield  {title} {\bibinfo {title} {Measurement-induced continuous time crystals},\ }\href {https://doi.org/10.1103/PhysRevLett.130.150401} {\bibfield  {journal} {\bibinfo  {journal} {Phys. Rev. Lett.}\ }\textbf {\bibinfo {volume} {130}},\ \bibinfo {pages} {150401} (\bibinfo {year} {2023}{\natexlab{a}})}\BibitemShut {NoStop}%
\bibitem [{\citenamefont {Solanki}\ \emph {et~al.}(2022)\citenamefont {Solanki}, \citenamefont {Jaseem}, \citenamefont {Hajdu\ifmmode~\check{s}\else \v{s}\fi{}ek},\ and\ \citenamefont {Vinjanampathy}}]{solanki2022role}%
  \BibitemOpen
  \bibfield  {author} {\bibinfo {author} {\bibfnamefont {P.}~\bibnamefont {Solanki}}, \bibinfo {author} {\bibfnamefont {N.}~\bibnamefont {Jaseem}}, \bibinfo {author} {\bibfnamefont {M.}~\bibnamefont {Hajdu\ifmmode~\check{s}\else \v{s}\fi{}ek}},\ and\ \bibinfo {author} {\bibfnamefont {S.}~\bibnamefont {Vinjanampathy}},\ }\bibfield  {title} {\bibinfo {title} {Role of coherence and degeneracies in quantum synchronization},\ }\href {https://doi.org/10.1103/PhysRevA.105.L020401} {\bibfield  {journal} {\bibinfo  {journal} {Phys. Rev. A}\ }\textbf {\bibinfo {volume} {105}},\ \bibinfo {pages} {L020401} (\bibinfo {year} {2022})}\BibitemShut {NoStop}%
\bibitem [{\citenamefont {Hajdu\ifmmode~\check{s}\else \v{s}\fi{}ek}\ \emph {et~al.}(2022)\citenamefont {Hajdu\ifmmode~\check{s}\else \v{s}\fi{}ek}, \citenamefont {Solanki}, \citenamefont {Fazio},\ and\ \citenamefont {Vinjanampathy}}]{seeding2022michal}%
  \BibitemOpen
  \bibfield  {author} {\bibinfo {author} {\bibfnamefont {M.}~\bibnamefont {Hajdu\ifmmode~\check{s}\else \v{s}\fi{}ek}}, \bibinfo {author} {\bibfnamefont {P.}~\bibnamefont {Solanki}}, \bibinfo {author} {\bibfnamefont {R.}~\bibnamefont {Fazio}},\ and\ \bibinfo {author} {\bibfnamefont {S.}~\bibnamefont {Vinjanampathy}},\ }\bibfield  {title} {\bibinfo {title} {Seeding crystallization in time},\ }\href {https://doi.org/10.1103/PhysRevLett.128.080603} {\bibfield  {journal} {\bibinfo  {journal} {Phys. Rev. Lett.}\ }\textbf {\bibinfo {volume} {128}},\ \bibinfo {pages} {080603} (\bibinfo {year} {2022})}\BibitemShut {NoStop}%
\bibitem [{\citenamefont {Hurtado-Guti\'errez}\ \emph {et~al.}(2020)\citenamefont {Hurtado-Guti\'errez}, \citenamefont {Carollo}, \citenamefont {P\'erez-Espigares},\ and\ \citenamefont {Hurtado}}]{hurtado2020raretimecrystal}%
  \BibitemOpen
  \bibfield  {author} {\bibinfo {author} {\bibfnamefont {R.}~\bibnamefont {Hurtado-Guti\'errez}}, \bibinfo {author} {\bibfnamefont {F.}~\bibnamefont {Carollo}}, \bibinfo {author} {\bibfnamefont {C.}~\bibnamefont {P\'erez-Espigares}},\ and\ \bibinfo {author} {\bibfnamefont {P.~I.}\ \bibnamefont {Hurtado}},\ }\bibfield  {title} {\bibinfo {title} {Building continuous time crystals from rare events},\ }\href {https://doi.org/10.1103/PhysRevLett.125.160601} {\bibfield  {journal} {\bibinfo  {journal} {Phys. Rev. Lett.}\ }\textbf {\bibinfo {volume} {125}},\ \bibinfo {pages} {160601} (\bibinfo {year} {2020})}\BibitemShut {NoStop}%
\bibitem [{\citenamefont {Alaeian}\ and\ \citenamefont {Bu{\v{c}}a}(2022)}]{alaeian2022exact}%
  \BibitemOpen
  \bibfield  {author} {\bibinfo {author} {\bibfnamefont {H.}~\bibnamefont {Alaeian}}\ and\ \bibinfo {author} {\bibfnamefont {B.}~\bibnamefont {Bu{\v{c}}a}},\ }\bibfield  {title} {\bibinfo {title} {Exact multistability and dissipative time crystals in interacting fermionic lattices},\ }\href {https://doi.org/10.1038/s42005-022-01090-z} {\bibfield  {journal} {\bibinfo  {journal} {Commun. Phys.}\ }\textbf {\bibinfo {volume} {5}},\ \bibinfo {pages} {318} (\bibinfo {year} {2022})}\BibitemShut {NoStop}%
\bibitem [{\citenamefont {Buonaiuto}\ \emph {et~al.}(2021)\citenamefont {Buonaiuto}, \citenamefont {Carollo}, \citenamefont {Olmos},\ and\ \citenamefont {Lesanovsky}}]{PhysRevLett.127.133601}%
  \BibitemOpen
  \bibfield  {author} {\bibinfo {author} {\bibfnamefont {G.}~\bibnamefont {Buonaiuto}}, \bibinfo {author} {\bibfnamefont {F.}~\bibnamefont {Carollo}}, \bibinfo {author} {\bibfnamefont {B.}~\bibnamefont {Olmos}},\ and\ \bibinfo {author} {\bibfnamefont {I.}~\bibnamefont {Lesanovsky}},\ }\bibfield  {title} {\bibinfo {title} {Dynamical phases and quantum correlations in an emitter-waveguide system with feedback},\ }\href {https://doi.org/10.1103/PhysRevLett.127.133601} {\bibfield  {journal} {\bibinfo  {journal} {Phys. Rev. Lett.}\ }\textbf {\bibinfo {volume} {127}},\ \bibinfo {pages} {133601} (\bibinfo {year} {2021})}\BibitemShut {NoStop}%
\bibitem [{\citenamefont {Nakanishi}\ and\ \citenamefont {Sasamoto}(2023)}]{PhysRevA.107.L010201}%
  \BibitemOpen
  \bibfield  {author} {\bibinfo {author} {\bibfnamefont {Y.}~\bibnamefont {Nakanishi}}\ and\ \bibinfo {author} {\bibfnamefont {T.}~\bibnamefont {Sasamoto}},\ }\bibfield  {title} {\bibinfo {title} {Dissipative time crystals originating from parity-time symmetry},\ }\href {https://doi.org/10.1103/PhysRevA.107.L010201} {\bibfield  {journal} {\bibinfo  {journal} {Phys. Rev. A}\ }\textbf {\bibinfo {volume} {107}},\ \bibinfo {pages} {L010201} (\bibinfo {year} {2023})}\BibitemShut {NoStop}%
\bibitem [{\citenamefont {Iemini}\ \emph {et~al.}(2024)\citenamefont {Iemini}, \citenamefont {Chang},\ and\ \citenamefont {Marino}}]{iemini2024sectors}%
  \BibitemOpen
  \bibfield  {author} {\bibinfo {author} {\bibfnamefont {F.}~\bibnamefont {Iemini}}, \bibinfo {author} {\bibfnamefont {D.}~\bibnamefont {Chang}},\ and\ \bibinfo {author} {\bibfnamefont {J.}~\bibnamefont {Marino}},\ }\bibfield  {title} {\bibinfo {title} {Dynamics of inhomogeneous spin ensembles with all-to-all interactions: Breaking permutational invariance},\ }\href {https://doi.org/10.1103/PhysRevA.109.032204} {\bibfield  {journal} {\bibinfo  {journal} {Phys. Rev. A}\ }\textbf {\bibinfo {volume} {109}},\ \bibinfo {pages} {032204} (\bibinfo {year} {2024})}\BibitemShut {NoStop}%
\bibitem [{\citenamefont {Cabot}\ \emph {et~al.}(2024{\natexlab{a}})\citenamefont {Cabot}, \citenamefont {Giorgi},\ and\ \citenamefont {Zambrini}}]{cabot2023nonequilibrium}%
  \BibitemOpen
  \bibfield  {author} {\bibinfo {author} {\bibfnamefont {A.}~\bibnamefont {Cabot}}, \bibinfo {author} {\bibfnamefont {G.~L.}\ \bibnamefont {Giorgi}},\ and\ \bibinfo {author} {\bibfnamefont {R.}~\bibnamefont {Zambrini}},\ }\bibfield  {title} {\bibinfo {title} {Nonequilibrium transition between dissipative time crystals},\ }\href {https://doi.org/10.1103/PRXQuantum.5.030325} {\bibfield  {journal} {\bibinfo  {journal} {PRX Quantum}\ }\textbf {\bibinfo {volume} {5}},\ \bibinfo {pages} {030325} (\bibinfo {year} {2024}{\natexlab{a}})}\BibitemShut {NoStop}%
\bibitem [{\citenamefont {Mattes}\ \emph {et~al.}(2023)\citenamefont {Mattes}, \citenamefont {Lesanovsky},\ and\ \citenamefont {Carollo}}]{mattes2023entangled}%
  \BibitemOpen
  \bibfield  {author} {\bibinfo {author} {\bibfnamefont {R.}~\bibnamefont {Mattes}}, \bibinfo {author} {\bibfnamefont {I.}~\bibnamefont {Lesanovsky}},\ and\ \bibinfo {author} {\bibfnamefont {F.}~\bibnamefont {Carollo}},\ }\bibfield  {title} {\bibinfo {title} {Entangled time-crystal phase in an open quantum light-matter system},\ }\href {https://doi.org/10.1103/PhysRevA.108.062216} {\bibfield  {journal} {\bibinfo  {journal} {Phys. Rev. A}\ }\textbf {\bibinfo {volume} {108}},\ \bibinfo {pages} {062216} (\bibinfo {year} {2023})}\BibitemShut {NoStop}%
\bibitem [{\citenamefont {Solanki}\ \emph {et~al.}(2024)\citenamefont {Solanki}, \citenamefont {Krishna}, \citenamefont {Hajdu\ifmmode~\check{s}\else \v{s}\fi{}ek}, \citenamefont {Bruder},\ and\ \citenamefont {Vinjanampathy}}]{solanki2024exotic}%
  \BibitemOpen
  \bibfield  {author} {\bibinfo {author} {\bibfnamefont {P.}~\bibnamefont {Solanki}}, \bibinfo {author} {\bibfnamefont {M.}~\bibnamefont {Krishna}}, \bibinfo {author} {\bibfnamefont {M.}~\bibnamefont {Hajdu\ifmmode~\check{s}\else \v{s}\fi{}ek}}, \bibinfo {author} {\bibfnamefont {C.}~\bibnamefont {Bruder}},\ and\ \bibinfo {author} {\bibfnamefont {S.}~\bibnamefont {Vinjanampathy}},\ }\bibfield  {title} {\bibinfo {title} {Exotic synchronization in continuous time crystals outside the symmetric subspace},\ }\href {https://doi.org/10.1103/PhysRevLett.133.260403} {\bibfield  {journal} {\bibinfo  {journal} {Phys. Rev. Lett.}\ }\textbf {\bibinfo {volume} {133}},\ \bibinfo {pages} {260403} (\bibinfo {year} {2024})}\BibitemShut {NoStop}%
\bibitem [{\citenamefont {Mukherjee}\ \emph {et~al.}(2024)\citenamefont {Mukherjee}, \citenamefont {Ibrahim}, \citenamefont {Hajdu\ifmmode~\check{s}\else \v{s}\fi{}ek},\ and\ \citenamefont {Vinjanampathy}}]{Mukherjee2024correlations}%
  \BibitemOpen
  \bibfield  {author} {\bibinfo {author} {\bibfnamefont {A.}~\bibnamefont {Mukherjee}}, \bibinfo {author} {\bibfnamefont {Y.}~\bibnamefont {Ibrahim}}, \bibinfo {author} {\bibfnamefont {M.}~\bibnamefont {Hajdu\ifmmode~\check{s}\else \v{s}\fi{}ek}},\ and\ \bibinfo {author} {\bibfnamefont {S.}~\bibnamefont {Vinjanampathy}},\ }\bibfield  {title} {\bibinfo {title} {Symmetries and correlations in continuous time crystals},\ }\href {https://doi.org/10.1103/PhysRevA.110.012220} {\bibfield  {journal} {\bibinfo  {journal} {Phys. Rev. A}\ }\textbf {\bibinfo {volume} {110}},\ \bibinfo {pages} {012220} (\bibinfo {year} {2024})}\BibitemShut {NoStop}%
\bibitem [{\citenamefont {Carollo}\ \emph {et~al.}(2024)\citenamefont {Carollo}, \citenamefont {Lesanovsky}, \citenamefont {Antezza},\ and\ \citenamefont {Chiara}}]{igor_thermodynamics}%
  \BibitemOpen
  \bibfield  {author} {\bibinfo {author} {\bibfnamefont {F.}~\bibnamefont {Carollo}}, \bibinfo {author} {\bibfnamefont {I.}~\bibnamefont {Lesanovsky}}, \bibinfo {author} {\bibfnamefont {M.}~\bibnamefont {Antezza}},\ and\ \bibinfo {author} {\bibfnamefont {G.~D.}\ \bibnamefont {Chiara}},\ }\bibfield  {title} {\bibinfo {title} {Quantum thermodynamics of boundary time-crystals},\ }\href {https://doi.org/10.1088/2058-9565/ad3f42} {\bibfield  {journal} {\bibinfo  {journal} {Quantum Sci. Technol.}\ }\textbf {\bibinfo {volume} {9}},\ \bibinfo {pages} {035024} (\bibinfo {year} {2024})}\BibitemShut {NoStop}%
\bibitem [{\citenamefont {Paulino}\ \emph {et~al.}(2024)\citenamefont {Paulino}, \citenamefont {Cabot}, \citenamefont {Chiara}, \citenamefont {Antezza}, \citenamefont {Lesanovsky},\ and\ \citenamefont {Carollo}}]{paulino2024thermodynamicscoupledtimecrystals}%
  \BibitemOpen
  \bibfield  {author} {\bibinfo {author} {\bibfnamefont {P.~J.}\ \bibnamefont {Paulino}}, \bibinfo {author} {\bibfnamefont {A.}~\bibnamefont {Cabot}}, \bibinfo {author} {\bibfnamefont {G.~D.}\ \bibnamefont {Chiara}}, \bibinfo {author} {\bibfnamefont {M.}~\bibnamefont {Antezza}}, \bibinfo {author} {\bibfnamefont {I.}~\bibnamefont {Lesanovsky}},\ and\ \bibinfo {author} {\bibfnamefont {F.}~\bibnamefont {Carollo}},\ }\href {https://arxiv.org/abs/2411.04836} {\bibinfo {title} {Thermodynamics of coupled time crystals with an application to energy storage}} (\bibinfo {year} {2024}),\ \Eprint {https://arxiv.org/abs/2411.04836} {arXiv:2411.04836 [quant-ph]} \BibitemShut {NoStop}%
\bibitem [{\citenamefont {Paulino}\ \emph {et~al.}(2023)\citenamefont {Paulino}, \citenamefont {Lesanovsky},\ and\ \citenamefont {Carollo}}]{PhysRevA.108.023516}%
  \BibitemOpen
  \bibfield  {author} {\bibinfo {author} {\bibfnamefont {P.~J.}\ \bibnamefont {Paulino}}, \bibinfo {author} {\bibfnamefont {I.}~\bibnamefont {Lesanovsky}},\ and\ \bibinfo {author} {\bibfnamefont {F.}~\bibnamefont {Carollo}},\ }\bibfield  {title} {\bibinfo {title} {Nonequilibrium thermodynamics and power generation in open quantum optomechanical systems},\ }\href {https://doi.org/10.1103/PhysRevA.108.023516} {\bibfield  {journal} {\bibinfo  {journal} {Phys. Rev. A}\ }\textbf {\bibinfo {volume} {108}},\ \bibinfo {pages} {023516} (\bibinfo {year} {2023})}\BibitemShut {NoStop}%
\bibitem [{\citenamefont {Solanki}\ and\ \citenamefont {Minganti}(2024)}]{solanki2024chaos}%
  \BibitemOpen
  \bibfield  {author} {\bibinfo {author} {\bibfnamefont {P.}~\bibnamefont {Solanki}}\ and\ \bibinfo {author} {\bibfnamefont {F.}~\bibnamefont {Minganti}},\ }\href {https://arxiv.org/abs/2411.07297} {\bibinfo {title} {Chaos in time: A dissipative continuous quasi time crystals}} (\bibinfo {year} {2024}),\ \Eprint {https://arxiv.org/abs/2411.07297} {arXiv:2411.07297 [quant-ph]} \BibitemShut {NoStop}%
\bibitem [{\citenamefont {Nadolny}\ \emph {et~al.}(2024)\citenamefont {Nadolny}, \citenamefont {Bruder},\ and\ \citenamefont {Brunelli}}]{nadolny2024nonreciprocal}%
  \BibitemOpen
  \bibfield  {author} {\bibinfo {author} {\bibfnamefont {T.}~\bibnamefont {Nadolny}}, \bibinfo {author} {\bibfnamefont {C.}~\bibnamefont {Bruder}},\ and\ \bibinfo {author} {\bibfnamefont {M.}~\bibnamefont {Brunelli}},\ }\href {https://arxiv.org/abs/2406.03357} {\bibinfo {title} {Nonreciprocal synchronization of active quantum spins}} (\bibinfo {year} {2024}),\ \Eprint {https://arxiv.org/abs/2406.03357} {arXiv:2406.03357 [quant-ph]} \BibitemShut {NoStop}%
\bibitem [{\citenamefont {Cabot}\ \emph {et~al.}(2024{\natexlab{b}})\citenamefont {Cabot}, \citenamefont {Carollo},\ and\ \citenamefont {Lesanovsky}}]{cabot2023continuous}%
  \BibitemOpen
  \bibfield  {author} {\bibinfo {author} {\bibfnamefont {A.}~\bibnamefont {Cabot}}, \bibinfo {author} {\bibfnamefont {F.}~\bibnamefont {Carollo}},\ and\ \bibinfo {author} {\bibfnamefont {I.}~\bibnamefont {Lesanovsky}},\ }\bibfield  {title} {\bibinfo {title} {Continuous sensing and parameter estimation with the boundary time crystal},\ }\href {https://doi.org/10.1103/PhysRevLett.132.050801} {\bibfield  {journal} {\bibinfo  {journal} {Phys. Rev. Lett.}\ }\textbf {\bibinfo {volume} {132}},\ \bibinfo {pages} {050801} (\bibinfo {year} {2024}{\natexlab{b}})}\BibitemShut {NoStop}%
\bibitem [{\citenamefont {Montenegro}\ \emph {et~al.}(2023)\citenamefont {Montenegro}, \citenamefont {Genoni}, \citenamefont {Bayat},\ and\ \citenamefont {Paris}}]{montenegro2023quantum}%
  \BibitemOpen
  \bibfield  {author} {\bibinfo {author} {\bibfnamefont {V.}~\bibnamefont {Montenegro}}, \bibinfo {author} {\bibfnamefont {M.~G.}\ \bibnamefont {Genoni}}, \bibinfo {author} {\bibfnamefont {A.}~\bibnamefont {Bayat}},\ and\ \bibinfo {author} {\bibfnamefont {M.~G.}\ \bibnamefont {Paris}},\ }\bibfield  {title} {\bibinfo {title} {Quantum metrology with boundary time crystals},\ }\href {https://doi.org/10.1038/s42005-023-01423-6} {\bibfield  {journal} {\bibinfo  {journal} {Commun. Phys.}\ }\textbf {\bibinfo {volume} {6}},\ \bibinfo {pages} {304} (\bibinfo {year} {2023})}\BibitemShut {NoStop}%
\bibitem [{\citenamefont {Gribben}\ \emph {et~al.}(2024)\citenamefont {Gribben}, \citenamefont {Sanpera}, \citenamefont {Fazio}, \citenamefont {Marino},\ and\ \citenamefont {Iemini}}]{gribben2024quantum}%
  \BibitemOpen
  \bibfield  {author} {\bibinfo {author} {\bibfnamefont {D.}~\bibnamefont {Gribben}}, \bibinfo {author} {\bibfnamefont {A.}~\bibnamefont {Sanpera}}, \bibinfo {author} {\bibfnamefont {R.}~\bibnamefont {Fazio}}, \bibinfo {author} {\bibfnamefont {J.}~\bibnamefont {Marino}},\ and\ \bibinfo {author} {\bibfnamefont {F.}~\bibnamefont {Iemini}},\ }\href {https://arxiv.org/abs/2406.06273} {\bibinfo {title} {Quantum enhancements and entropic constraints to boundary time crystals as sensors of {AC} fields}} (\bibinfo {year} {2024}),\ \Eprint {https://arxiv.org/abs/2406.06273} {arXiv:2406.06273 [quant-ph]} \BibitemShut {NoStop}%
\bibitem [{sup()}]{supp}%
  \BibitemOpen
  \href@noop {} {}\bibinfo {note} {See supplemental material for details.}\BibitemShut {Stop}%
\bibitem [{\citenamefont {Garbe}\ \emph {et~al.}(2024)\citenamefont {Garbe}, \citenamefont {Minoguchi}, \citenamefont {Huber},\ and\ \citenamefont {Rabl}}]{garbe2024}%
  \BibitemOpen
  \bibfield  {author} {\bibinfo {author} {\bibfnamefont {L.}~\bibnamefont {Garbe}}, \bibinfo {author} {\bibfnamefont {Y.}~\bibnamefont {Minoguchi}}, \bibinfo {author} {\bibfnamefont {J.}~\bibnamefont {Huber}},\ and\ \bibinfo {author} {\bibfnamefont {P.}~\bibnamefont {Rabl}},\ }\bibfield  {title} {\bibinfo {title} {{The bosonic skin effect: Boundary condensation in asymmetric transport}},\ }\href {https://doi.org/10.21468/SciPostPhys.16.1.029} {\bibfield  {journal} {\bibinfo  {journal} {SciPost Phys.}\ }\textbf {\bibinfo {volume} {16}},\ \bibinfo {pages} {029} (\bibinfo {year} {2024})}\BibitemShut {NoStop}%
\bibitem [{\citenamefont {Bu{\v{c}}a}\ and\ \citenamefont {Prosen}(2012)}]{Buca2012strong}%
  \BibitemOpen
  \bibfield  {author} {\bibinfo {author} {\bibfnamefont {B.}~\bibnamefont {Bu{\v{c}}a}}\ and\ \bibinfo {author} {\bibfnamefont {T.}~\bibnamefont {Prosen}},\ }\bibfield  {title} {\bibinfo {title} {A note on symmetry reductions of the {L}indblad equation: transport in constrained open spin chains},\ }\href {https://doi.org/10.1088/1367-2630/14/7/073007} {\bibfield  {journal} {\bibinfo  {journal} {New J. Phys.}\ }\textbf {\bibinfo {volume} {14}},\ \bibinfo {pages} {073007} (\bibinfo {year} {2012})}\BibitemShut {NoStop}%
\bibitem [{\citenamefont {Albert}\ and\ \citenamefont {Jiang}(2014)}]{Albert2014strong}%
  \BibitemOpen
  \bibfield  {author} {\bibinfo {author} {\bibfnamefont {V.~V.}\ \bibnamefont {Albert}}\ and\ \bibinfo {author} {\bibfnamefont {L.}~\bibnamefont {Jiang}},\ }\bibfield  {title} {\bibinfo {title} {Symmetries and conserved quantities in {L}indblad master equations},\ }\href {https://doi.org/10.1103/PhysRevA.89.022118} {\bibfield  {journal} {\bibinfo  {journal} {Phys. Rev. A}\ }\textbf {\bibinfo {volume} {89}},\ \bibinfo {pages} {022118} (\bibinfo {year} {2014})}\BibitemShut {NoStop}%
\bibitem [{\citenamefont {Kac}\ \emph {et~al.}(1963)\citenamefont {Kac}, \citenamefont {Uhlenbeck},\ and\ \citenamefont {Hemmer}}]{kac1963van}%
  \BibitemOpen
  \bibfield  {author} {\bibinfo {author} {\bibfnamefont {M.}~\bibnamefont {Kac}}, \bibinfo {author} {\bibfnamefont {G.}~\bibnamefont {Uhlenbeck}},\ and\ \bibinfo {author} {\bibfnamefont {P.}~\bibnamefont {Hemmer}},\ }\bibfield  {title} {\bibinfo {title} {On the van der {W}aals theory of the vapor-liquid equilibrium. {I}. {D}iscussion of a one-dimensional model},\ }\href {https://doi.org/10.1063/1.1703946} {\bibfield  {journal} {\bibinfo  {journal} {Journal of Mathematical Physics}\ }\textbf {\bibinfo {volume} {4}},\ \bibinfo {pages} {216} (\bibinfo {year} {1963})}\BibitemShut {NoStop}%
\bibitem [{\citenamefont {Nakanishi}\ \emph {et~al.}(2024)\citenamefont {Nakanishi}, \citenamefont {Hanai},\ and\ \citenamefont {Sasamoto}}]{nakanishi2024CTCpt}%
  \BibitemOpen
  \bibfield  {author} {\bibinfo {author} {\bibfnamefont {Y.}~\bibnamefont {Nakanishi}}, \bibinfo {author} {\bibfnamefont {R.}~\bibnamefont {Hanai}},\ and\ \bibinfo {author} {\bibfnamefont {T.}~\bibnamefont {Sasamoto}},\ }\href {https://arxiv.org/abs/2406.09018} {\bibinfo {title} {Continuous time crystals as a {PT} symmetric state and the emergence of critical exceptional points}} (\bibinfo {year} {2024}),\ \Eprint {https://arxiv.org/abs/2406.09018} {arXiv:2406.09018 [quant-ph]} \BibitemShut {NoStop}%
\bibitem [{\citenamefont {Bender}(2019)}]{bender2019pt}%
  \BibitemOpen
  \bibfield  {author} {\bibinfo {author} {\bibfnamefont {C.~M.}\ \bibnamefont {Bender}},\ }\href@noop {} {\emph {\bibinfo {title} {PT symmetry: In quantum and classical physics}}}\ (\bibinfo  {publisher} {World Scientific},\ \bibinfo {year} {2019})\BibitemShut {NoStop}%
\bibitem [{\citenamefont {Prosen}(2012)}]{Prosen_2012}%
  \BibitemOpen
  \bibfield  {author} {\bibinfo {author} {\bibfnamefont {T.}~\bibnamefont {Prosen}},\ }\bibfield  {title} {\bibinfo {title} {{PT}-symmetric quantum {Liouvillean} dynamics},\ }\href {https://doi.org/10.1103/physrevlett.109.090404} {\bibfield  {journal} {\bibinfo  {journal} {Phys. Rev. Lett.}\ }\textbf {\bibinfo {volume} {109}},\ \bibinfo {pages} {090404} (\bibinfo {year} {2012})}\BibitemShut {NoStop}%
\bibitem [{\citenamefont {Huber}\ \emph {et~al.}(2020)\citenamefont {Huber}, \citenamefont {Kirton}, \citenamefont {Rotter},\ and\ \citenamefont {Rabl}}]{huber2020}%
  \BibitemOpen
  \bibfield  {author} {\bibinfo {author} {\bibfnamefont {J.}~\bibnamefont {Huber}}, \bibinfo {author} {\bibfnamefont {P.}~\bibnamefont {Kirton}}, \bibinfo {author} {\bibfnamefont {S.}~\bibnamefont {Rotter}},\ and\ \bibinfo {author} {\bibfnamefont {P.}~\bibnamefont {Rabl}},\ }\bibfield  {title} {\bibinfo {title} {{Emergence of PT-symmetry breaking in open quantum systems}},\ }\href {https://doi.org/10.21468/SciPostPhys.9.4.052} {\bibfield  {journal} {\bibinfo  {journal} {SciPost Phys.}\ }\textbf {\bibinfo {volume} {9}},\ \bibinfo {pages} {052} (\bibinfo {year} {2020})}\BibitemShut {NoStop}%
\bibitem [{\citenamefont {Nakanishi}\ and\ \citenamefont {Sasamoto}(2022{\natexlab{a}})}]{nakanishi2022}%
  \BibitemOpen
  \bibfield  {author} {\bibinfo {author} {\bibfnamefont {Y.}~\bibnamefont {Nakanishi}}\ and\ \bibinfo {author} {\bibfnamefont {T.}~\bibnamefont {Sasamoto}},\ }\bibfield  {title} {\bibinfo {title} {$\mathcal{PT}$ phase transition in open quantum systems with {L}indblad dynamics},\ }\href {https://doi.org/10.1103/PhysRevA.105.022219} {\bibfield  {journal} {\bibinfo  {journal} {Phys. Rev. A}\ }\textbf {\bibinfo {volume} {105}},\ \bibinfo {pages} {022219} (\bibinfo {year} {2022}{\natexlab{a}})}\BibitemShut {NoStop}%
\bibitem [{\citenamefont {Nakanishi}\ and\ \citenamefont {Sasamoto}(2022{\natexlab{b}})}]{nakanishi2022dissipative}%
  \BibitemOpen
  \bibfield  {author} {\bibinfo {author} {\bibfnamefont {Y.}~\bibnamefont {Nakanishi}}\ and\ \bibinfo {author} {\bibfnamefont {T.}~\bibnamefont {Sasamoto}},\ }\bibfield  {title} {\bibinfo {title} {Dissipative time crystals originated from parity-time symmetry},\ }\href {https://arxiv.org/abs/2203.06672} {\bibfield  {journal} {\bibinfo  {journal} {arXiv:2203.06672}\ } (\bibinfo {year} {2022}{\natexlab{b}})}\BibitemShut {NoStop}%
\bibitem [{\citenamefont {Roccati}\ \emph {et~al.}(2021)\citenamefont {Roccati}, \citenamefont {Lorenzo}, \citenamefont {Palma}, \citenamefont {Landi}, \citenamefont {Brunelli},\ and\ \citenamefont {Ciccarello}}]{Roccati_2021}%
  \BibitemOpen
  \bibfield  {author} {\bibinfo {author} {\bibfnamefont {F.}~\bibnamefont {Roccati}}, \bibinfo {author} {\bibfnamefont {S.}~\bibnamefont {Lorenzo}}, \bibinfo {author} {\bibfnamefont {G.~M.}\ \bibnamefont {Palma}}, \bibinfo {author} {\bibfnamefont {G.~T.}\ \bibnamefont {Landi}}, \bibinfo {author} {\bibfnamefont {M.}~\bibnamefont {Brunelli}},\ and\ \bibinfo {author} {\bibfnamefont {F.}~\bibnamefont {Ciccarello}},\ }\bibfield  {title} {\bibinfo {title} {Quantum correlations in $\mathcal{PT}$-symmetric systems},\ }\href {https://doi.org/10.1088/2058-9565/abcfcc} {\bibfield  {journal} {\bibinfo  {journal} {Quantum Sci. Technol.}\ }\textbf {\bibinfo {volume} {6}},\ \bibinfo {pages} {025005} (\bibinfo {year} {2021})}\BibitemShut {NoStop}%
\bibitem [{\citenamefont {Arovas}\ and\ \citenamefont {Auerbach}(1988)}]{schwingertrasnformation1988}%
  \BibitemOpen
  \bibfield  {author} {\bibinfo {author} {\bibfnamefont {D.~P.}\ \bibnamefont {Arovas}}\ and\ \bibinfo {author} {\bibfnamefont {A.}~\bibnamefont {Auerbach}},\ }\bibfield  {title} {\bibinfo {title} {Functional integral theories of low-dimensional quantum {H}eisenberg models},\ }\href {https://doi.org/10.1103/PhysRevB.38.316} {\bibfield  {journal} {\bibinfo  {journal} {Phys. Rev. B}\ }\textbf {\bibinfo {volume} {38}},\ \bibinfo {pages} {316} (\bibinfo {year} {1988})}\BibitemShut {NoStop}%
\bibitem [{\citenamefont {Breuer}\ and\ \citenamefont {Petruccione}(2002)}]{breuer2002theory}%
  \BibitemOpen
  \bibfield  {author} {\bibinfo {author} {\bibfnamefont {H.-P.}\ \bibnamefont {Breuer}}\ and\ \bibinfo {author} {\bibfnamefont {F.}~\bibnamefont {Petruccione}},\ }\href {https://doi.org/10.1093/acprof:oso/9780199213900.001.0001} {\emph {\bibinfo {title} {The theory of open quantum systems}}}\ (\bibinfo  {publisher} {Oxford University Press on Demand},\ \bibinfo {year} {2002})\BibitemShut {NoStop}%
\bibitem [{\citenamefont {Gonzalez-Ballestero}(2024)}]{gonzalez2024tutorial}%
  \BibitemOpen
  \bibfield  {author} {\bibinfo {author} {\bibfnamefont {C.}~\bibnamefont {Gonzalez-Ballestero}},\ }\bibfield  {title} {\bibinfo {title} {Tutorial: projector approach to master equations for open quantum systems},\ }\href {https://doi.org/10.22331/q-2024-08-29-1454} {\bibfield  {journal} {\bibinfo  {journal} {Quantum}\ }\textbf {\bibinfo {volume} {8}},\ \bibinfo {pages} {1454} (\bibinfo {year} {2024})}\BibitemShut {NoStop}%
\bibitem [{\citenamefont {Krishna}\ \emph {et~al.}(2023{\natexlab{b}})\citenamefont {Krishna}, \citenamefont {Solanki},\ and\ \citenamefont {Vinjanampathy}}]{krishna2023select}%
  \BibitemOpen
  \bibfield  {author} {\bibinfo {author} {\bibfnamefont {M.}~\bibnamefont {Krishna}}, \bibinfo {author} {\bibfnamefont {P.}~\bibnamefont {Solanki}},\ and\ \bibinfo {author} {\bibfnamefont {S.}~\bibnamefont {Vinjanampathy}},\ }\bibfield  {title} {\bibinfo {title} {Select topics in open quantum systems},\ }\href {https://link.springer.com/article/10.1007/s41745-022-00338-5} {\bibfield  {journal} {\bibinfo  {journal} {J. Indian Inst. Sci.}\ }\textbf {\bibinfo {volume} {103}},\ \bibinfo {pages} {513} (\bibinfo {year} {2023}{\natexlab{b}})}\BibitemShut {NoStop}%
\bibitem [{\citenamefont {Lesanovsky}\ and\ \citenamefont {Garrahan}(2013)}]{kinetic2013igor}%
  \BibitemOpen
  \bibfield  {author} {\bibinfo {author} {\bibfnamefont {I.}~\bibnamefont {Lesanovsky}}\ and\ \bibinfo {author} {\bibfnamefont {J.~P.}\ \bibnamefont {Garrahan}},\ }\bibfield  {title} {\bibinfo {title} {Kinetic constraints, hierarchical relaxation, and onset of glassiness in strongly interacting and dissipative {R}ydberg gases},\ }\href {https://doi.org/10.1103/PhysRevLett.111.215305} {\bibfield  {journal} {\bibinfo  {journal} {Phys. Rev. Lett.}\ }\textbf {\bibinfo {volume} {111}},\ \bibinfo {pages} {215305} (\bibinfo {year} {2013})}\BibitemShut {NoStop}%
\bibitem [{\citenamefont {Carr}\ \emph {et~al.}(2013)\citenamefont {Carr}, \citenamefont {Ritter}, \citenamefont {Wade}, \citenamefont {Adams},\ and\ \citenamefont {Weatherill}}]{PhysRevLett.111.113901}%
  \BibitemOpen
  \bibfield  {author} {\bibinfo {author} {\bibfnamefont {C.}~\bibnamefont {Carr}}, \bibinfo {author} {\bibfnamefont {R.}~\bibnamefont {Ritter}}, \bibinfo {author} {\bibfnamefont {C.~G.}\ \bibnamefont {Wade}}, \bibinfo {author} {\bibfnamefont {C.~S.}\ \bibnamefont {Adams}},\ and\ \bibinfo {author} {\bibfnamefont {K.~J.}\ \bibnamefont {Weatherill}},\ }\bibfield  {title} {\bibinfo {title} {Nonequilibrium phase transition in a dilute {R}ydberg ensemble},\ }\href {https://doi.org/10.1103/PhysRevLett.111.113901} {\bibfield  {journal} {\bibinfo  {journal} {Phys. Rev. Lett.}\ }\textbf {\bibinfo {volume} {111}},\ \bibinfo {pages} {113901} (\bibinfo {year} {2013})}\BibitemShut {NoStop}%
\bibitem [{\citenamefont {Marcuzzi}\ \emph {et~al.}(2014)\citenamefont {Marcuzzi}, \citenamefont {Levi}, \citenamefont {Diehl}, \citenamefont {Garrahan},\ and\ \citenamefont {Lesanovsky}}]{PhysRevLett.113.210401}%
  \BibitemOpen
  \bibfield  {author} {\bibinfo {author} {\bibfnamefont {M.}~\bibnamefont {Marcuzzi}}, \bibinfo {author} {\bibfnamefont {E.}~\bibnamefont {Levi}}, \bibinfo {author} {\bibfnamefont {S.}~\bibnamefont {Diehl}}, \bibinfo {author} {\bibfnamefont {J.~P.}\ \bibnamefont {Garrahan}},\ and\ \bibinfo {author} {\bibfnamefont {I.}~\bibnamefont {Lesanovsky}},\ }\bibfield  {title} {\bibinfo {title} {Universal nonequilibrium properties of dissipative {R}ydberg gases},\ }\href {https://doi.org/10.1103/PhysRevLett.113.210401} {\bibfield  {journal} {\bibinfo  {journal} {Phys. Rev. Lett.}\ }\textbf {\bibinfo {volume} {113}},\ \bibinfo {pages} {210401} (\bibinfo {year} {2014})}\BibitemShut {NoStop}%
\bibitem [{\citenamefont {Boneberg}\ \emph {et~al.}(2022)\citenamefont {Boneberg}, \citenamefont {Lesanovsky},\ and\ \citenamefont {Carollo}}]{boneberg2022quantum}%
  \BibitemOpen
  \bibfield  {author} {\bibinfo {author} {\bibfnamefont {M.}~\bibnamefont {Boneberg}}, \bibinfo {author} {\bibfnamefont {I.}~\bibnamefont {Lesanovsky}},\ and\ \bibinfo {author} {\bibfnamefont {F.}~\bibnamefont {Carollo}},\ }\bibfield  {title} {\bibinfo {title} {Quantum fluctuations and correlations in open quantum {D}icke models},\ }\href {https://doi.org/10.1103/PhysRevA.106.012212} {\bibfield  {journal} {\bibinfo  {journal} {Phys. Rev. A}\ }\textbf {\bibinfo {volume} {106}},\ \bibinfo {pages} {012212} (\bibinfo {year} {2022})}\BibitemShut {NoStop}%
\bibitem [{\citenamefont {Dutta}\ \emph {et~al.}(2024)\citenamefont {Dutta}, \citenamefont {Zhang},\ and\ \citenamefont {Haque}}]{dutta2024quantumoriginlimitcycles}%
  \BibitemOpen
  \bibfield  {author} {\bibinfo {author} {\bibfnamefont {S.}~\bibnamefont {Dutta}}, \bibinfo {author} {\bibfnamefont {S.}~\bibnamefont {Zhang}},\ and\ \bibinfo {author} {\bibfnamefont {M.}~\bibnamefont {Haque}},\ }\href {https://arxiv.org/abs/2405.08866} {\bibinfo {title} {On the quantum origin of limit cycles, fixed points, and critical slowing down}} (\bibinfo {year} {2024}),\ \Eprint {https://arxiv.org/abs/2405.08866} {arXiv:2405.08866 [quant-ph]} \BibitemShut {NoStop}%
\bibitem [{\citenamefont {Olivares}(2012)}]{olivares2012quantum}%
  \BibitemOpen
  \bibfield  {author} {\bibinfo {author} {\bibfnamefont {S.}~\bibnamefont {Olivares}},\ }\bibfield  {title} {\bibinfo {title} {Quantum optics in the phase space: A tutorial on {G}aussian states},\ }\href {https://doi.org/10.1140/epjst/e2012-01532-4} {\bibfield  {journal} {\bibinfo  {journal} {Eur. Phys. J. Spec. Top.}\ }\textbf {\bibinfo {volume} {203}},\ \bibinfo {pages} {3} (\bibinfo {year} {2012})}\BibitemShut {NoStop}%
\bibitem [{\citenamefont {Adesso}\ \emph {et~al.}(2014)\citenamefont {Adesso}, \citenamefont {Ragy},\ and\ \citenamefont {Lee}}]{adesso2014continuous}%
  \BibitemOpen
  \bibfield  {author} {\bibinfo {author} {\bibfnamefont {G.}~\bibnamefont {Adesso}}, \bibinfo {author} {\bibfnamefont {S.}~\bibnamefont {Ragy}},\ and\ \bibinfo {author} {\bibfnamefont {A.~R.}\ \bibnamefont {Lee}},\ }\bibfield  {title} {\bibinfo {title} {Continuous variable quantum information: {G}aussian states and beyond},\ }\href {https://doi.org/10.1142/S1230161214400010} {\bibfield  {journal} {\bibinfo  {journal} {Open Syst. Inf. Dyn.}\ }\textbf {\bibinfo {volume} {21}},\ \bibinfo {pages} {1440001} (\bibinfo {year} {2014})}\BibitemShut {NoStop}%
\bibitem [{\citenamefont {Adesso}\ \emph {et~al.}(2005)\citenamefont {Adesso}, \citenamefont {Serafini},\ and\ \citenamefont {Illuminati}}]{adesso2005entanglement}%
  \BibitemOpen
  \bibfield  {author} {\bibinfo {author} {\bibfnamefont {G.}~\bibnamefont {Adesso}}, \bibinfo {author} {\bibfnamefont {A.}~\bibnamefont {Serafini}},\ and\ \bibinfo {author} {\bibfnamefont {F.}~\bibnamefont {Illuminati}},\ }\bibfield  {title} {\bibinfo {title} {Entanglement, purity, and information entropies in continuous variable systems},\ }\href {https://doi.org/10.1007/s11080-005-5730-2} {\bibfield  {journal} {\bibinfo  {journal} {Open Syst. Inf. Dyn.}\ }\textbf {\bibinfo {volume} {12}},\ \bibinfo {pages} {189} (\bibinfo {year} {2005})}\BibitemShut {NoStop}%
\end{thebibliography}
\end{document}